\documentclass[longauth]{aa}
\usepackage{graphicx}
\usepackage{natbib}
\defcitealias{2025olah}{Paper~I}
\usepackage{subcaption}
\usepackage{txfonts}
\usepackage{amsmath}
\usepackage{multicol}
\usepackage{stfloats}
\usepackage{color}
\usepackage{xcolor}
\usepackage{yfonts}
\usepackage{ulem}
\usepackage[usenames,dvipsnames]{xcolor}
\usepackage{hyperref}
\hypersetup{
    colorlinks=true,
    citecolor = blue,
    linkcolor=blue
            }

\begin{document} 
   \title{Starspots on eclipsing giant stars}
   \subtitle{II.~TESS continuous viewing zone giant V521\,Dra
   }

  \author{A.~Haris\inst{1}
\and
B.~Seli
   \inst{2,3,4}
\and
K.~Ol\'ah
   \inst{2,3}
\and
S.~Rappaport
   \inst{5}
\and
J.~S.~Thomsen
\inst{6}
\and
M.~Turkki
\inst{7,8,9}
\and
T.~Hackman
\inst{1}
\and
M.~Tuomi
\inst{1}
\and
K.~Yakut
\inst{10,11}
\and
A.~Casasbuenas
\inst{12}
\and
M.~A.~Díaz
\inst{8}
\and
J.~H.~Terwel
\inst{13}
\and
B.~N.~Hauptmann
\inst{14}
\and
R.~Gagliano
\inst{15}
\and
T.~L.~Jacobs
\inst{16}
\and
M.~H.~Kristiansen
\inst{17}
\and
H.~M.~Schwengeler
\inst{18}
\and
M.~Omohundro
\inst{18}
\and
I.~Terentev
\inst{18}
\and
B.~Powell
\inst{19}
\and
V.~Kostov
\inst{19,20}
\and
Zs.~Kővári
   \inst{2,3}
}
   \institute{
   Department of Physics, P.O.~Box 64, 00014 University of Helsinki, Finland\\
   \email{andras.haris-kiss@helsinki.fi}
   \and
   Konkoly Observatory, HUN-REN Research Centre for Astronomy and Earth Sciences, Konkoly Thege Mikl\'os \'ut 15-17., H-1121, Budapest, Hungary
   \and 
   CSFK, MTA Centre of Excellence, Budapest, Konkoly Thege Mikl\'os út 15-17., H-1121, Hungary
   \and
   E\"otv\"os University, Department of Astronomy, Pf. 32, 1518 Budapest, Hungary
   \and
   Department of Physics, Kavli Institute for Astrophysics and Space Research, Massachusetts Institute of Technology, Cambridge, MA 02139, USA
   \and
   Dipartimento di Fisica e Astronomia, Università di Bologna, Via Zamboni, 33, Bologna, 40126, Italia
   \and
   Finnish Centre for Astronomy with ESO (FINCA), University of Turku, FI-20014 Turku, Finland
   \and
   Department of Physics and Astronomy, University of Turku, Vesilinnantie 5, Turku FI-20014, Finland
   \and
   Aalto University Mesähovi Radio Observatory, Metsähovintie 114, 02540 Kylmälä, Finland
   \and
   Department of Astronomy and Space Sciences, Faculty of Science, Ege University, 35100, İzmir, Türkiye
   \and
   Institute of Astronomy, The Observatories, Madingley Road, Cambridge CB3 OHA, UK
   \and
   Universidad de La Laguna, Dpto.~Astrofísica, 38206 La Laguna, Tenerife, Spain
   \and
   School of Physics, Trinity College Dublin, The University of Dublin, Dublin 2, Ireland
   \and
   Institut de Física d'Altes Energies, Campus UAB, Facultat Ciencies Nord, 08193 Bellaterra, Barcelona, Spain
   \and
   Amateur Astronomer, Glendale, AZ 85308, USA
   \and
   Amateur Astronomer, 1319 Missouri City, TX 77459, USA
   \and
   Brorfelde Observatory, Observator Gyldenkernes Vej 7, DK-4340 Tølløse, Denmark
   \and
   Citizen Scientist, c/o Zooniverse, Department of Physics, University of Oxford, Denys Wilkinson Building, Keble Road, Oxford OX1 3RH, UK
   \and
   NASA Goddard Space Flight Center, 8800 Greenbelt Road, Greenbelt, MD 20771, USA
   \and
   SETI Institute, 189 Bernardo Avenue, Suite 200, Mountain View, CA 94043, USA
}             
   \date{Received ...; accepted ...}

\abstract
{Eclipsing binaries are exceptional laboratories for stellar astrophysics. By analysing the modulations in their eclipse light curves, we can probe their surface inhomogeneities. The eclipse-mapping technique enables us to scan starspots in detail, using the transit of a companion star across the surface of a spotted star. When multiple consecutive eclipses are observed, this method may also reveal the evolution of spot properties and provide evidence for stellar differential rotation.}
{We study the magnetic activity of V521\,Dra, a tidally locked eclipsing binary with a red giant primary ($M_{\rm A} = 1.70 \pm 0.10~M_\odot$) and a main-sequence secondary ($M_{\rm B} = 0.74 \pm 0.04~M_\odot$), both of K spectral class, on a $4.27$~day orbit. This is an exceptional system with $\sim$5.3~years of TESS 2\,min cadence observations, including over two hundred eclipses showing spot-crossing events. This offers a unique opportunity to study magnetic activity in uninterrupted blocks over several years with very high photometric precision. We aim to constrain the evolution of the spot pattern of V521\,Dra~A.}
{We used two methods to study the spot activity of the primary: eclipse mapping and analytic spot modelling. We analysed the stellar properties of the system with a set of methods, including combined modelling of the eclipse light curves, radial velocity data, spectral energy distribution analysis, and forward modelling of the whole orbital light curve.}
{The star shows strong spot-induced modulations both in the eclipse and out-of-eclipse parts of the light curve. We identified $540$ spot eclipse signals in $203$ primary eclipses. Eclipse mapping reveals the presence of up to four spots. The spot longitudes drift continuously in time, indicating differential rotation. Spots in the eclipse-mapping data show differing angular velocities in their drift, implying a near-surface shear in the convective zone of the primary. Using time series spectroscopy and the spectral energy distribution, we constrained the astrophysical parameters of the binary system.}
{}

   \maketitle
   \nolinenumbers

\section{Introduction}

The magnetic activity of low-mass stars can be observed in every stage of their evolution, from the pre-main sequence through the main sequence to the red giant branch (RGB). The origin of magnetic fields in evolved stars has been extensively discussed by \citet{charbonnel2017}. More recently, \citet{ortiz2026arXiv} presented a theoretical study that also provides a comprehensive overview of the subject and the relevant literature.
The initial mass and binarity (or multiplicity) have important roles in the evolution of a star. These same factors also influence the star’s convective structure and rotation rate, which in turn shape its magnetic activity. This connection has been demonstrated both for evolved stars, through Doppler imaging of the binary system XX\,Tri by \citet{strassmeier2024}, and for rapidly rotating main-sequence binaries by \citet{yu2025}.
The proper characterisation of the stellar properties and magnetic activity  of red giant (RG) stars has been challenging until recently, since it requires the continuous monitoring of these stars, which was unavailable before space-based photometry. In addition, giants are low in number due to their rapid evolution after the main sequence. However, when a giant star is part of an eclipsing binary, reliable absolute parameters, such as masses and radii, can be obtained \citep{gaulme2016, brogaard2018, thomsen2025}.

Binarity helps to develop and maintain strong magnetic activity due to rapid rotation through synchronisation \citep{2025CoSka..55c.182K}. The precise characterisation of the binary system helps us to understand the counteractions between the different forces (e.g.~gravitational, magnetic, and dynamic) and their connections. As the orbital separation decreases due to tidal interactions and magnetic braking, proximity effects become increasingly significant.

Based on Kepler photometry, \citet{2014ApJ...785....5G} investigated the activity of red giants in binary systems, showing that the magnetic field suppresses the oscillation of giant stars in eclipsing binaries.
The difference in the magnetic activity between single giant stars and members of close binaries reveals that the tidal locking may result in larger magnetic fields and more effective suppression of the oscillations \citep{2020A&A...639A..63G}. For more complete overviews of the topic, we refer readers to to \citet{gehan2022, gehan2024}.

The original aim of this series of papers was to find giant stars in  eclipsing binaries whose rotational brightness variation due to magnetic activity grossly exceeds the ellipsoidal light variations, so we could study the light variation more clearly, with minimum possible interference from the companion star. 
To find such binaries is challenging, since the number of giants in binary systems is generally about $50\%$ for stars around a few solar masses \citep[][Fig.\,2]{2021ApJS..253...32W}. In addition, the number of eclipsing binaries among binary systems is very low, about one percent, or slightly more in the case of close binaries \citep{2011AJ....141...83P}. In \citet[hereafter \citetalias{2025olah}]{2025olah}, after checking the observed Transiting Exoplanet Survey Satellite \citep[TESS,][]{ricker2015} sectors for about five years, we listed 29 close binaries with primaries showing large amplitude rotational modulation (i.e.~due to spots) relative to the ellipsoidal variations and, as a pilot project, analysed three of them. Details of the target selection are given in Sect.\,2 of \citetalias{2025olah}.

Before space-based photometry, eclipse mapping was explored using both an analytical approach by \citet{budding1988} and a surface integration method by \citet{collier1997}. Due to the sparse photometric coverage, starspots have been studied less extensively in binaries with giant primaries. Modelling full rotational light curves suffers from degeneracies between the radii and effective temperatures of starspots, and it provides only approximate constraints on spot latitudes. These limitations are described in detail by \citet[][see their Introduction and references]{luger2021}, based on the same formalism that we use in this paper. Eclipse mapping, which allows for a more reliable characterisation of starspots, albeit over only a limited area of the stellar surface, was generally not feasible with ground-based data due to the limited accuracy of the observations. In addition, the eclipses in these systems often last longer than 12~hours, making continuous observations with ground-based telescopes difficult.

High-precision space photometry, which has been widely available since the  Kepler \citep{borucki2010} and TESS telescopes, allows us to observe subtle features, such as shallow eclipses of binaries and low amplitude rotational modulation caused by spots. \citetalias{2025olah} briefly reviews the main results of eclipse mapping since its first application by \citet{silva2003}. For this method the continuous viewing zones (CVZs) of TESS are particularly valuable. These regions, covering approximately $5\%$ of the sky, are monitored continuously over extended periods, enabling the study of temporal changes in stellar activity. We found only four systems that fulfil our selection requirements in the CVZs of TESS, and analysed three of those in the pilot project. This paper presents refinements to the approach of eclipse mapping and modelling the full light curve, in addition to the thorough investigation of the eclipsing binary system V521\,Dra.

V521\,Dra (TIC~441804674) is an RS\,CVn variable, and it is the best observed of the $29$ eclipsing close binary systems catalogued in \citetalias{2025olah}. The quality and amount of available TESS data make this system an exceptional target for eclipse mapping. The first panel of Table\,\ref{table:1} lists the observable parameters of the system, known prior to this analysis. The system consists of a red giant primary and a main-sequence secondary, and it can be considered a rather close binary in the sample with non-negligible proximity effects. 
As suggested by our preliminary analysis, the two components have a significant flux difference, implying that any contribution from the secondary’s activity is likely negligible.
Our work presents the first dedicated and comprehensive analysis of V521\,Dra.

We applied a combination of methods to investigate the spot activity of the primary component. By comparing the spot signals originating from inside and outside the eclipsed region, we also sought to estimate the star's surface differential rotation. Overall, our goal was to characterise the properties and evolution of the starspots as comprehensively as possible in order to improve our understanding of surface magnetic activity, and modelling thereof, in red giants.

The structure of the paper is as follows. In Sect.\,\ref{sect:observations}, we provide a summary of the observations collected for this study. Sect.\,\ref{sect:spot_analysis_methods} presents the light curve analysis using eclipse mapping and full light curve modelling and the results regarding spot activity, while Sect.\,\ref{sect:stellar_properties} details the results from the analysis of the system’s stellar properties through the analysis of the radial velocity curve of the two components, the forward modelling of the whole orbital light curve, and spectral energy distribution analysis.  In Sect.\,\ref{sect:discussion}, we discuss our results in relation to the stellar evolution of the binary system, as well as the surface rotation of the red giant.

\begin{table}[t!]
\caption{System parameters.}
\label{table:1}
\centering
\begin{tabular}{l c}
\hline\hline
Parameter & Value \\ 
\hline
    Right ascension & 17:44:10.91 $^{(1)}$ \\
    Declination & +72:03:02.80 $^{(1)}$ \\
    Parallax (mas) & $3.10 \pm 0.01$ $^{(1)}$ \\
    Distance $d$~(pc) & $323 \pm 1$ $^{(1)}$ \\
    $B$-band magnitude & $11.17 \pm 0.06$ $^{(2)}$ \\
    $V$-band magnitude & $10.27 \pm 0.04$ $^{(2)}$ \\
    Luminosity $L~(L_{\odot})$ & $12.491$ $^{(3)}$ \\ 
    TESS magnitude $m_{\rm TESS}$ & $9.027 \pm 0.011$ $^{(4)}$ \\ \hline 
    Photometry & \\ \hline 
    Orbital period $P_{\rm orb}$ (d) & $4.270337 \pm 0.000004$ $^{(5)}$ \\
    Rotational period (A) $\overline{P}_{\rm rot}$ (d) & $4.257 \pm 0.002$ $^{(5)}$ \\
    Epoch $T_0$ (BJD) & $2\,458\,685.524 \pm 0.001$ $^{(5)}$ \\ \hline
    Eclipse mapping & \\ \hline 
    Flux ratio $k_f$ & $55.8 \pm 0.4$ $^{(5)}$ \\
    Radius ratio $k_r$ & $7.24 \pm 0.01$ $^{(5)}$ \\
    Inclination $i$~(deg) & $89.6^{+0.3}_{-0.4}$ $^{(5)}$ \\
    Orbital separation $a/R_{\rm A}$ & $2.42 \pm 0.01$ $^{(5)}$ \\
    Limb darkening $u_{1,\rm{A}}$ & $0.62 \pm 0.02$ $^{(5)}$ \\
    Limb darkening $u_{2,\rm{A}}$ & $-0.11 \pm 0.04 $ $^{(5)}$ \\
    Eclipse length $T_{14}$ (h) & $15.98 \pm 0.01$ $^{(5)}$ \\
    Scanned latitude range & $(-7.0_{-0.6}^{+1.0}\,^\circ, 8.9_{-0.6}^{+1.0}\,^\circ)$ $^{(5)}$ \\
    \hline
    Spectroscopy & \\       \hline 
    Semi amplitude $K_{\rm A}$ (km\,s$^{-1}$) & $54 \pm 1$ $^{(5)}$ \\
    Semi amplitude $K_{\rm B}$ (km\,s$^{-1}$) & $123 \pm 3$ $^{(5)}$ \\
    Eccentricity $e$ & $0.02 \pm 0.01$ $^{(5)}$ \\ 
    Argument of periastron $\omega_{\rm A}$ & $3 \pm 2$ $^{(5)}$ \\ 
    Mean anomaly $M_{\rm 0,A}$ & $3 \pm 1$ $^{(5)}$ \\ 
    Reference velocity $\gamma$ (km\,s$^{-1}$) & $-25.1 \pm 0.7$ $^{(5)}$ \\ 
    RV noise $\sigma_{\rm RV,A}$ (km\,s$^{-1}$) & $2 \pm 1$  $^{(5)}$ \\ 
    RV noise $\sigma_{\rm RV,B}$ (km\,s$^{-1}$) & $4 \pm 3$  $^{(5)}$ \\ 
    Chrom.~activity $\log R'_{\rm HK, A}$ & $-3.9 \pm 0.1$ $^{(5)}$ \\
    Surface gravity $\log g_{\rm A}$ (cgs) & $3.10 \pm 0.02$ $^{(5),~a}$\\
    Metallicity $[\mathrm{Fe/H}]$ & $0.07 \pm 0.09$ $^{(5)}$ \\
    Projected vel. $v \sin i$ (km\,s$^{-1}$) & $73 \pm 3$ $^{(5)}$ \\
\hline 
\end{tabular}
\tablefoot{
\tablefoottext{a}{Based on spectral energy distribution analysis.}
}
\begin{tablebib}
\noindent\raggedright \\ (1)~\citet{gaiadr3};
(2)~\citet{hog2000};
(3)~\citet{gaiadr2};
(4)~TESS Input Catalog v8.2 \citep{stassun2018};
(5)~This work.
\par
\end{tablebib}
\end{table}

\section{Observations \label{sect:observations}}

\subsection{TESS photometry}

Due to its high ecliptic latitude ($b = 84.33^\circ$), V521\,Dra was in the CVZ of TESS in its second year of operation (Sectors~14--26). The system was also surveyed by TESS in its fourth and seventh year, yielding 38~sectors of observations in total. All sectors contain 120\,s cadence observations.

We accessed the Mikulski Archive for Space Telescopes for the Simple Aperture Photometry (SAP) fluxes, reduced by the Science Processing Operations Center (SPOC) pipeline, using the \texttt{Lightkurve} Python package \citep{lightkurve2018}. We analysed the Target Pixel Files using \texttt{tpfplotter} \citep{aller2020} to assess potential contamination from nearby sources. We looked at the SAP aperture for V521\,Dra and compared it to the positions of nearby stars from Gaia~DR2 astrometry \citep{gaiadr2}. No nearby sources with a magnitude difference of $\Delta m \leq 6$ are located within the SAP aperture, suggesting minimal risk of photometric contamination.

\begin{figure}[t!]
    \includegraphics[width=\columnwidth]{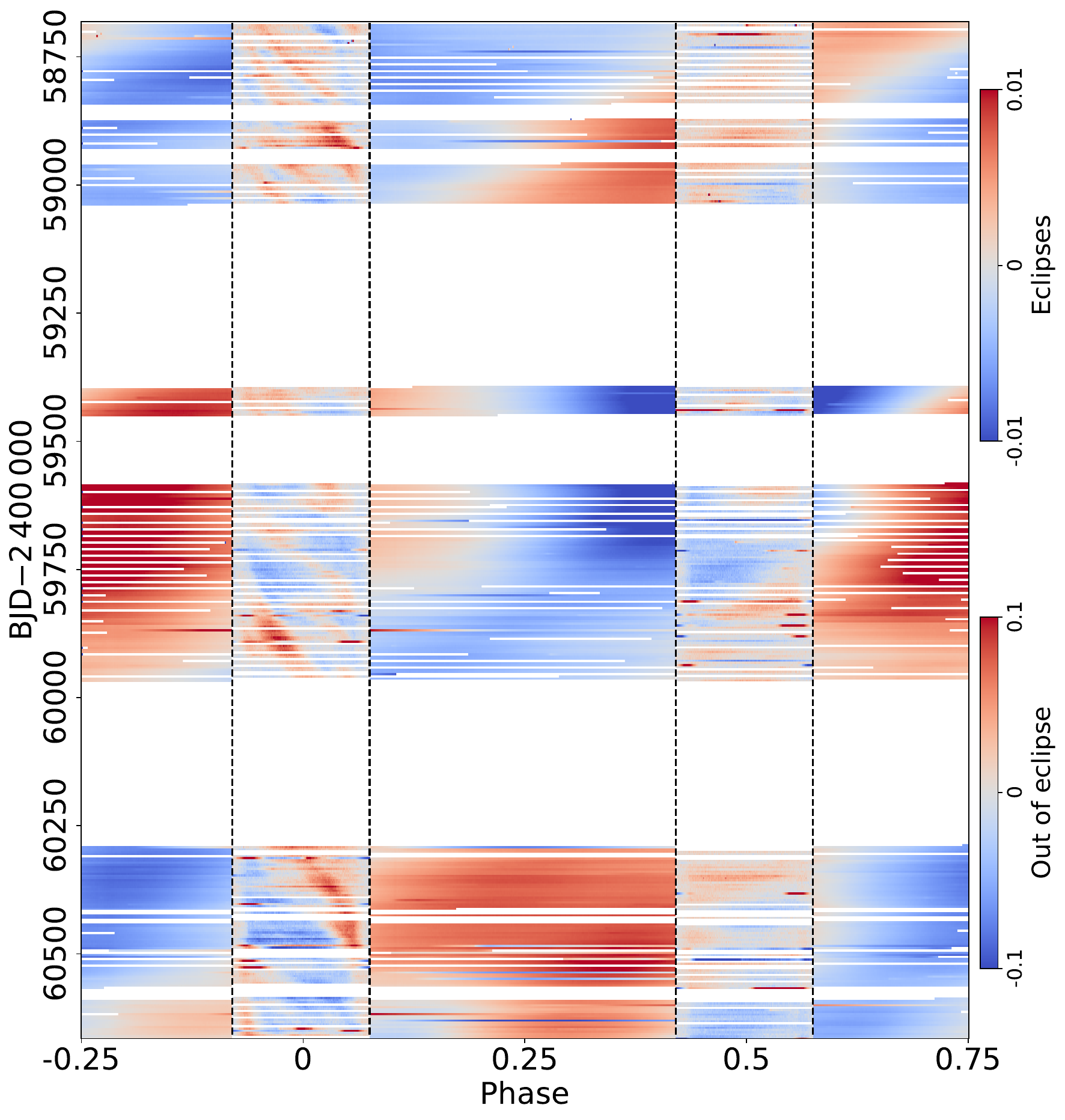}
      \caption{TESS light curve of V521\,Dra folded with the orbital period. To suppress the effect of binarity and enhance the spot modulation signal, a time-averaged phase curve is subtracted. The eclipses are shown between dashed lines. Primary eclipses, clearly indicating starspot crossings, are on the left, while secondary eclipses are on the right.}
      \label{full_map}
\end{figure}

We show the TESS light curves of V521\,Dra folded with the orbital period of the binary in Fig.\,\ref{full_map}. A time-averaged phase curve has been subtracted to amplify the spot induced variations. Both the in- and out-of-eclipse parts of the light curve show strong variations and continuous changes, implying strong spot activity and spot evolution. It can also be seen that while the in- and out-of-eclipse parts trace the same spots, the eclipses reveal more detailed information about the spots inside the eclipse chord. Just by visually inspecting Fig.\,\ref{full_map}, we can see that the secondary star crosses over one to three spots during each eclipse --- a  fact we discuss more formally in Sect.\,\ref{sect:spot_analysis_methods} in the context of model comparison.

\subsection{FIES spectroscopy \label{sec:spectroscopy}}

We used spectroscopic observations for radial velocity measurements, obtained with the high-resolution FIbre-fed Echelle Spectrograph \citep[FIES,][]{telting2014} at the $2.56$~m Nordic Optical Telescope (NOT) located at the Observatorio del Roque de los Muchachos on La Palma. The spectral resolution was $R \sim 46\,000$ and the integration time was 600\,s. \texttt{FIEStool} \citep{stempels2017} was used for data reduction, including \'echelle order extraction, wavelength calibration, and order merging. We obtained eight spectra, spread over $38$~days.

\section{Spot analysis and characterisation \label{sect:spot_analysis_methods}}

\subsection{Eclipse mapping \label{sec:transit_mapping}}

We determined the orbital period of the system $P_{\rm orb}$ and the epoch $T_0$ (the mid-time of the first primary eclipse, i.e.~when the secondary component passes in front of the primary) and extracted the eclipse light curves using a slightly modified version of the method described in \citetalias{2025olah}. Appendix\,\ref{app:eclipse_lcs_method} gives a detailed description of the extraction of the light curves. We report $P_{\rm orb}$ and $T_0$, along with the rotational period of the primary (see Sect.\,\ref{sec:prot_A}) in the second panel of Table\,\ref{table:1}.

We followed the method of \citetalias{2025olah} to characterise the modulations from spot transits in the eclipse light curves of V521\,Dra. This method fits a two-component light curve model to the eclipses and assesses the likelihood of the models using Bayesian selection criteria. The components of the light curve models are 1)~an unspotted baseline of the eclipse light curve from the analytic eclipse model of \citet{mandel2002}, scaled by the fractional brightnesses of the binary components, and 2)~a modulation in flux from a spot eclipse.
The binary system, and the unspotted baseline of the eclipse light curve are parametrised by the flux and radius ratios $k_f = F_{\rm A}/F_{\rm B}$ and $k_r = R_{\rm A}/R_{\rm B}$, orbital inclination $i$, scaled separation $a/R_{\rm A}$, orbital period $P_{\rm orb}$, and two quadratic limb darkening coefficients $(u_1, u_2)$ per star. 
A spot occultation is characterised by the maximum amplitude of the spot occultation $A$, the beginning and end of the maximum-amplitude phase $t_{\rm in}$ and $t_{\rm out}$, and the scale of the exponential decay during spot ingress and egress, $\sigma_{\rm spot}$.
For a detailed description of the eclipse model, see \citetalias{2025olah}.

Unlike in \citetalias{2025olah}, we fixed the eclipse baseline to avoid the difficulties related to possible astrophysical noise and the ambiguity  whether the baseline was physically found, or the code just detected a long spot eclipse, as well as to speed up computations. The treatment of the eclipse baseline and the sampling used for eclipse mapping are presented in detail in Appendix\,\ref{app:eclipse_baseline}.

Similarly to the primary component, the secondary is  also expected to be magnetically active with significant spot coverage. We tested whether variability caused by spots on the secondary could influence the eclipse baselines or bias the spot properties derived from eclipse mapping, especially since the eclipse duration is a significant fraction of the orbital period. Assuming an extreme spot configuration on the secondary (one spot with radius $r = 25^\circ$, contrast $c=0.5$), the change in eclipse baseline is far below the TESS noise level. We also simulated rotational spot modulation from the secondary and found that, although it could in principle introduce variations of up to $0.002$ in the system brightness in normalised units, these effects are largely removed by the normalisation procedure.

\begin{figure*}[t!]
\includegraphics[width=\textwidth]{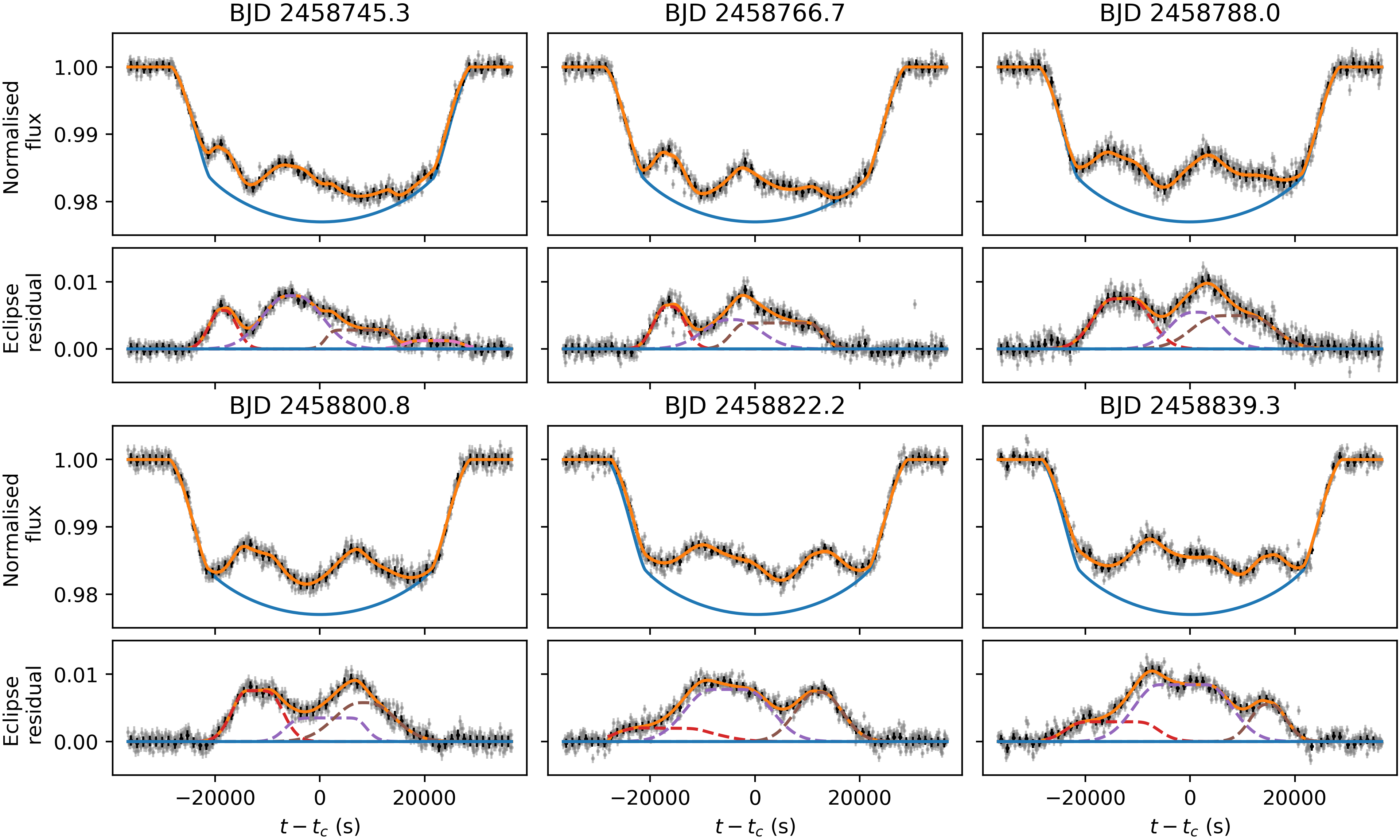}
    \caption{\label{fig:example_transits} Example primary eclipse light curves of V521\,Dra arranged as four rows of subplots forming two top–bottom pairs. In each pair, the upper panels show the original light curves in grey, while black dots show the mean of consecutive subsets of ten data points. The mid-time of each eclipse is given above the upper panels.
    Spot signals are shown in orange and the fixed eclipse baseline in blue. The lower panels show the corresponding residual light curves after the subtraction of the eclipse baseline, showing only the spot-induced modulations. Individual spot signals are plotted with dashed lines, each in a different colour.}
\end{figure*}

We analysed $203$ primary eclipses of V521\,Dra observed by TESS.
In Fig.\,\ref{fig:example_transits} we highlight a representative subset of these eclipses  showing several spot eclipse signals and signs of spot drifts (i.e.~changes in spot longitude in the orbital reference frame), possibly due to differential rotation.
The number of spot signals in each eclipse was determined by sampling models with $n_{\rm spots} = 0, ..., 4$ spots. For each model, the Bayesian Information Criterion (BIC) was calculated from the maximum likelihood obtained during posterior sampling and used to estimate the relative model probabilities. The corresponding Bayes factors were then used to assess the evidence in favour of models with different numbers of spots. We adopted the model-evidence scale of \citet{jeffreys1961} and \citet{kass1995}, according to which a Bayes factor of $B_{n+1,n}>150$ constitutes decisive evidence in favour of a model with $n+1$ spots over one with $n$ spots.

\begin{figure*}[t!]
\centering
\includegraphics[width=0.9\textwidth]{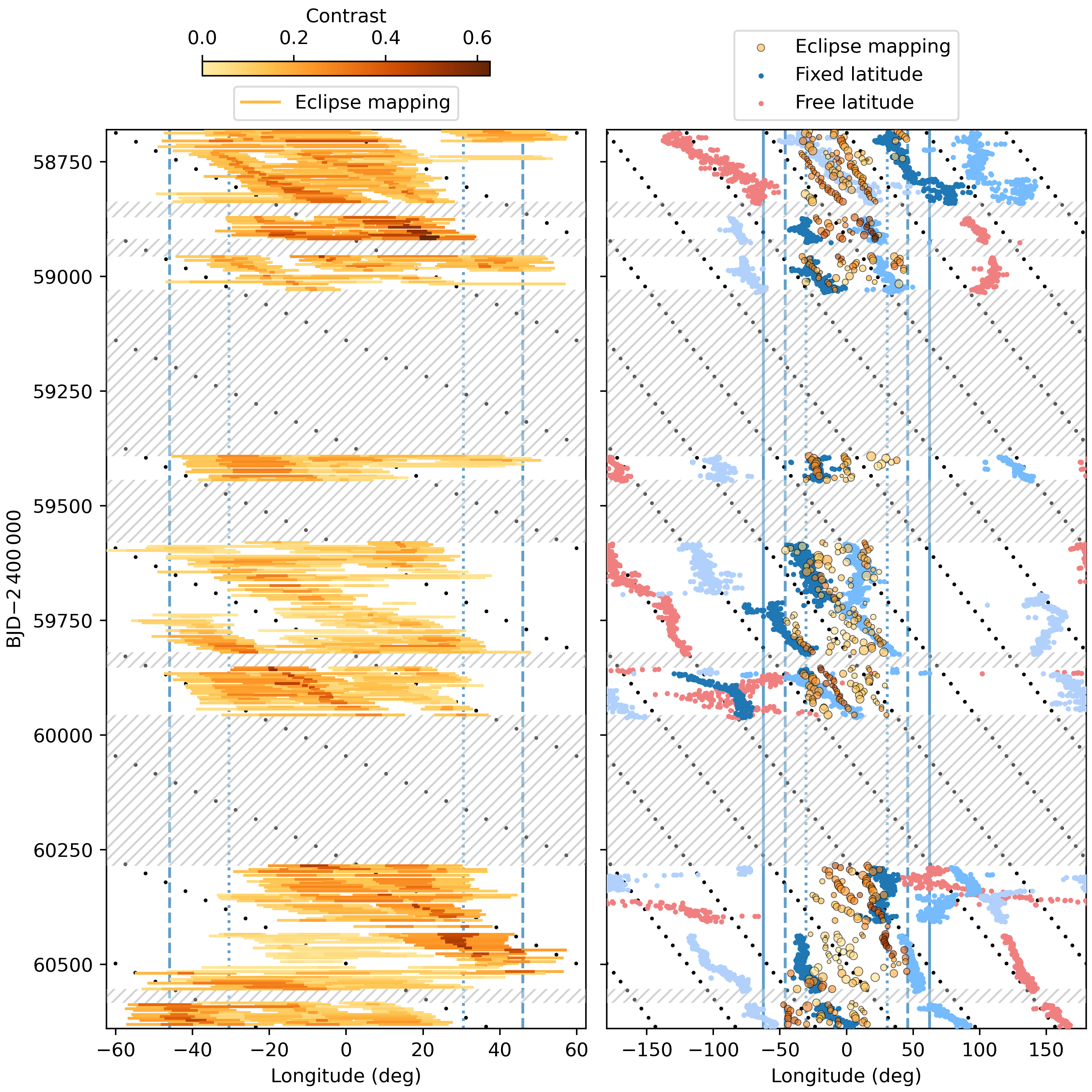}
      \caption{Left panel: Spot longitudes, sizes, and contrasts  from eclipse mapping over the full TESS time series, in the orbital reference frame. 
      Spots  are marked as grey lines with the shading of their inner, darker parts marking their contrasts.  Their inner part corresponds to the longitudes covered by the secondary between times $t_{\rm in}$ and $t_{\rm out}$, while the lighter  areas correspond to longitudes covered  between  $t_{\rm in} - \sigma_{\rm spot}$ and $t_{\rm out} + \sigma_{\rm spot}$.
       Longitudes within the dashed and dotted blue lines have a $>10\%$ and $>50\%$ detection probability by eclipse mapping, respectively.
      Right panel: Spot longitudes from full light curve modelling with four~spots, in the binary orbit reference frame. Spots marked with blue were fixed at $2^\circ$ latitude, while for spots marked with red, latitude was a free parameter. Each shading of blue corresponds to an individual spot signal. The longitude region available for eclipse mapping is within solid light  blue lines. The regions where spots can be detected with $>10\%$ and $> 50\%$ probability are marked as in the left  panel. Spots from eclipse mapping are marked with  grey dots. The angular velocity from the rotation of the stellar surface with average period $\overline{P}_{\rm rot}$ is marked with tilted black dotted lines on both  panels. Hatched bands mark missing TESS Sectors, when no observations were available.
      }
      \label{fig:tr_map}
\end{figure*}

We show all analysed primary eclipses along with the best-fit spot models in Fig.\,\ref{fig:all_eclipses} of Appendix\,\ref{app:eclipse_baseline}. We identified $540$ spot eclipse signals in the dataset. Spot radii, longitudes and contrasts were calculated for all eclipse signals. We disclose all spot properties in the Digital attachment. We present the longitude, size, and contrast of spots detected from eclipse mapping in the eclipse band over the whole TESS time series in detail in the left panel of Fig.\,\ref{fig:tr_map}. The right panel of Fig.\,\ref{fig:tr_map} compares these results with spot longitudes obtained from full light curve modelling (see Sect.\,\ref{full_lc}).

As shown in Fig.~\ref{fig:tr_map}, the spots drift in the orbital reference frame with an average angular velocity consistent with that expected from the stellar rotation period. The drift expected from the difference between the orbital and stellar rotation periods is marked with black dots in tilted lines. Individual spots, however, may drift faster or slower than this average trend, possibly owing to radial differential rotation and different anchoring depths. This possibility is discussed further in Sect.~\ref{sect:spot_evolution}.

Since the primary star is rapidly co-rotating with its companion, the longitude range  available for eclipse mapping is less than $(-90^\circ, 90^\circ)$, which would correspond to a non-rotating primary. Due to this synchronous rotation only longitudes $(-62.43 \pm 0.02^\circ, 62.43 \pm 0.02^\circ)$ are scanned by the secondary. In reality, this range is even smaller, as spots falling close to the edge of the eclipse have smaller amplitudes and are thus less likely to be confirmed by our method.

Due to projection effects, smaller and lower-contrast spots are increasingly difficult to detect towards the stellar limb. To quantify this effect, we simulated $10^5$ eclipse light curves with spots distributed across the scanned longitude range. The spot longitudes were drawn randomly, while their radii and contrasts were uniformly sampled between $0$ and $25^\circ$, and $0$ and $0.5$, respectively. The noise properties of the simulated light curves were matched to those of the TESS data.
The simulated spots were then grouped according to their angular distance from the stellar meridian, and for each bin we estimated the fraction of spots that would be detected based on our detection criteria. At an angular distance of $31^\circ$ from the primary's meridian, only $50\%$ of spots were recovered, while at $46^\circ$ the recovery rate dropped to $10\%$. Beyond $58^\circ$, no spots were recovered in our tests.

Spot radii estimated from eclipse mapping ranged up to $27 \pm 1$ degrees, while spot contrasts reached up to $0.63 \pm 0.01$, corresponding to effective temperature differences up to $\Delta T \sim 940$~K compared to the unspotted photosphere. We note that the effective temperature inferred for the eclipsed regions provides only an upper limit to the temperature of individual spots. This is because the contrast derived from the eclipse-mapping model corresponds to the entire region occulted by the secondary, rather than to individual spots \citep{haris2025}. The spot signal with the highest distance away from the central meridian of the primary was $47 \pm 2$ degrees. We did not find clear correlations between spot contrasts, radii and longitudes.

\subsection{Separating the orbital light curve from stellar activity}
\label{sec:disentangle}

\begin{figure}[t!]
    \includegraphics[width=\columnwidth]{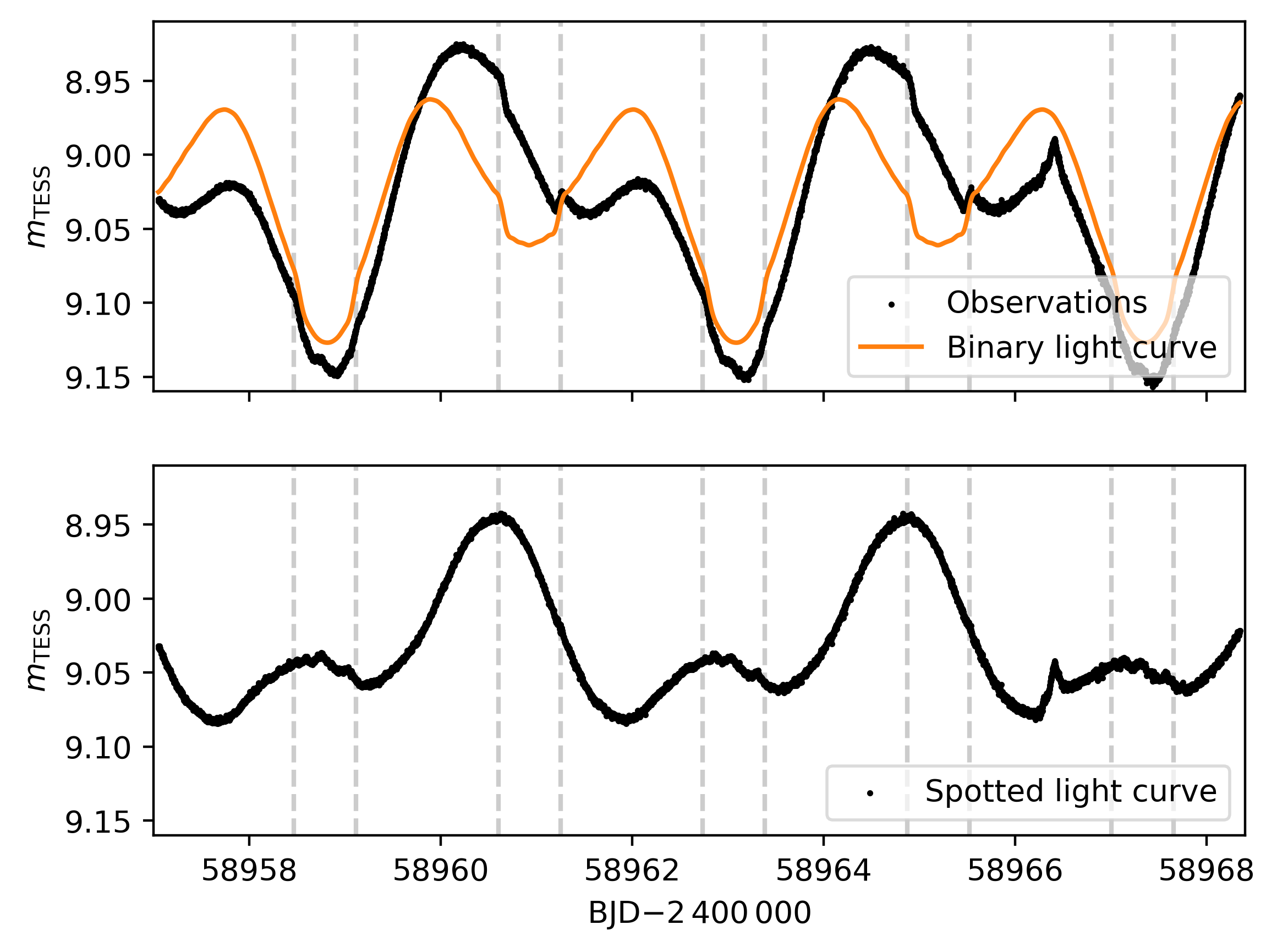}
      \caption{Example of separating the  brightness variation from spots from the  binary light curve. Top: original TESS flux measurements (black) and the binary light curve model (orange).
      Bottom: binary light curve model subtracted from the data, showing only the signatures of co-rotating surface features. Start and end times of eclipses are marked with vertical dashed lines.}
      \label{fig:example}
\end{figure}

Significant parts of the observed light variation of V521\,Dra ($\approx$$0.15$\,$m_{\rm TESS}$ amplitude) originate from the ellipsoidal light variations, due to the tidal distortion of the giant component, and from the eclipses. An additional difficulty comes from the fact that the mean of the SAP flux can vary significantly from one TESS sector to another.
For modelling the spots over whole stellar rotations, we need a light curve free from these effects. 
To obtain such a light curve, the observations from all sectors were median normalised, thereby eliminating the calibration differences from the different sectors.
The difference between the  observed and  binary light curves gives the true spotted variability (spot-only light curve) of the giant primary. As shown in Fig.\,\ref{fig:example}, the spotted light curve is markedly different from the eclipsing one. 

By modelling the full light curve, rather than only the eclipses, we also obtained information on spot coverage in regions of the primary’s surface that are not occulted by the secondary. We first removed the brightness variations caused by binarity in order to model the photometric variability produced by co-rotating spots on the surface of the primary star. We describe our method to obtain a phase curve of the binary, free from spot induced brightness variations, in Appendix\,\ref{app:separating_lc}.

A total of $35$~flares were identified in the SAP light curves. Appendix\,\ref{app:flares} briefly describes the flare statistics based on the TESS measurements. Fig.\,\ref{fig:flares} shows the phase folded light curve of the system with the flares marked in colour.

\subsection{Orbital period and rotational period of the primary star \label{sec:prot_A}}

The clear, spot-only light curve allowed us to find the average rotational period $\overline{P}_{\rm rot}$ of the giant primary originating purely from the spots from the full light curve as $4.257 \pm 0.002$~days using \textsc{mufran} \citep{2004ESASP.559..396C}. The errors  were calculated from the width of the periodogram peaks at 90\% height. This value significantly differs from the orbital period of 4.270337~days (Table\,\ref{table:1}).

The dataset was divided into nine segments separated by shorter (but more than one full rotation) or longer gaps in the data.
To verify the rotational period and obtain an independent estimate of its uncertainty, we also modelled each of the nine light curve segments with a quasi-periodic Gaussian process using the \texttt{RotationTerm} kernel implemented in \texttt{celerite2} \citep{celerite1}. The posterior distribution was sampled with \texttt{emcee} \citep{foreman2013} using wide Gaussian priors, and the period and its uncertainty were taken as the median and standard deviation of the posterior samples.

\begin{figure}[t]
    \includegraphics[width=\columnwidth]{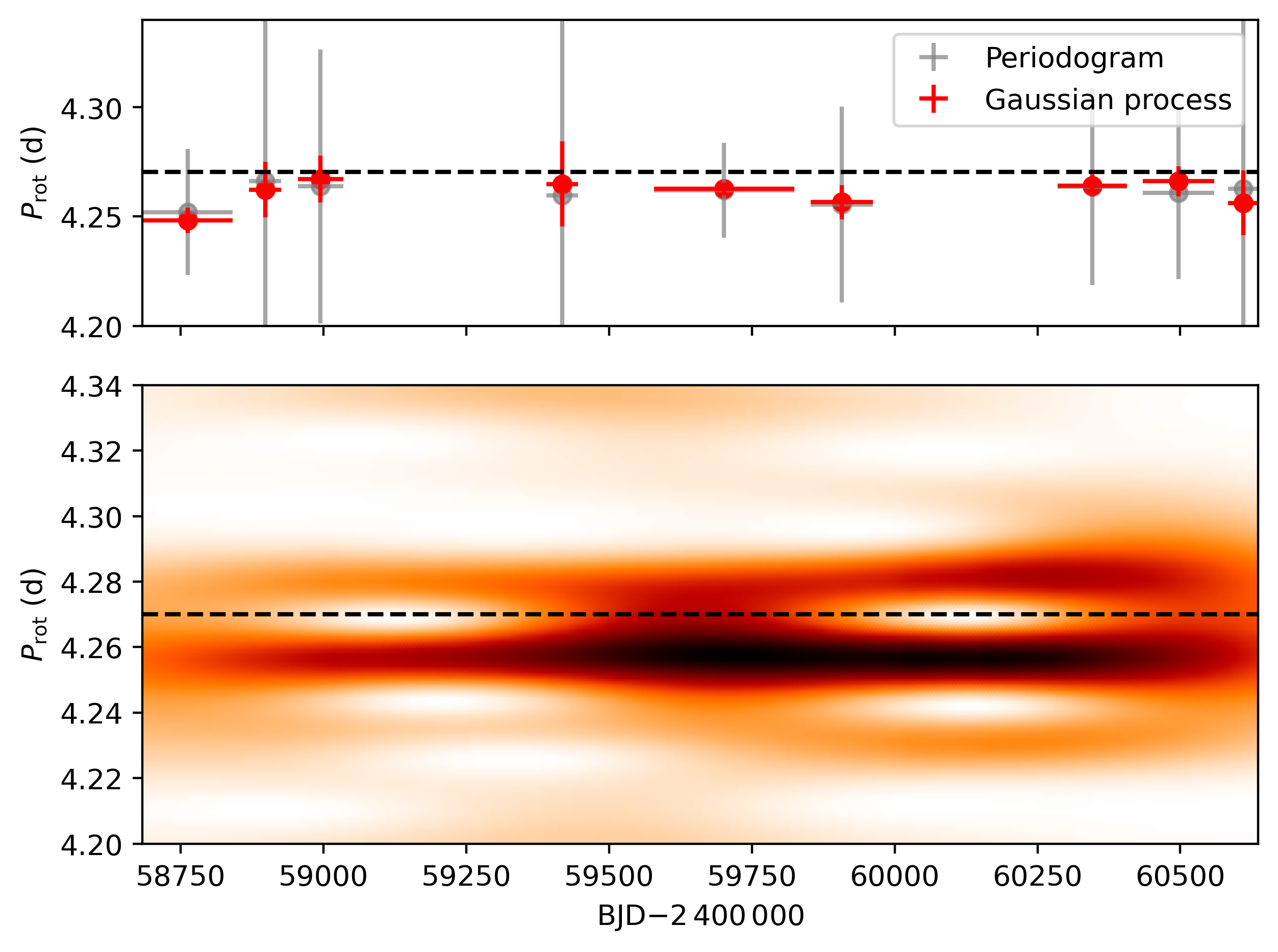}
    \caption{Top: Per-segment average rotational periods measured using periodogram analysis (\textsc{mufran}, \citealt{2004ESASP.559..396C}) and from the quasi-periodic Gaussian process. Bottom: Windowed Lomb-Scargle periodogram  \citep{2024JOSS....9.7118B} showing the temporal changes and structure of the rotational periods present in the data. The dashed line in both panels marks the orbital period of 4.270337 days.
    }
      \label{fig:rotper}
\end{figure}

Figure\,\ref{fig:rotper} shows the evolution of the rotation period of V521\,Dra~A in a complementary manner. The upper panel shows the rotation periods derived independently for the nine observation segments, along with the uncertainties derived from the periodogram and Gaussian process analyses. Although individual measurements show small deviations around the mean value, all estimates remain close to the orbital period, suggesting near-synchronous rotation. The lower panel presents the corresponding time-frequency diagram as a windowed Lomb--Scargle transform made with the \texttt{seismolab} package \citep{2024JOSS....9.7118B}, which shows a continuous and persistent power ridge centred near $P_{\rm orb}$ over the observation interval.

An interesting feature of both the segmental period determinations and the time-frequency diagram is that the dominant rotation signals
remain consistently shorter than the orbital period, suggesting that the active regions rotate somewhat faster than the synchronous rate,
consistent with weak surface differential rotation; for a detailed discussion of this, see Sect.\,\ref{sect:discussion_rotation}.
However, this feature may also indicate a small degree of supersynchronous rotation (i.e.~$P_{\rm rot}$ is shorter than $P_{\rm orb}$), or both effects (that is, differential rotation and supersynchronicity) can coexist.

\subsection{Full light curve modelling \label{full_lc}}

The spotted light curve of V521\,Dra was modelled in time-series mode of our own software; its performance is detailed in Sect.\,4.6 of \citetalias{2025olah}, and is not repeated here. Since we normalised all flux measurements by dividing them with the median of the TESS sector, all measurements were put on a common flux scale. This enabled choosing a single unspotted brightness of $8.90$ in TESS magnitude, which, in lack of information about the long-term behaviour of the star, is a lower limit.

\begin{figure}[t]
    \includegraphics[width=\columnwidth]{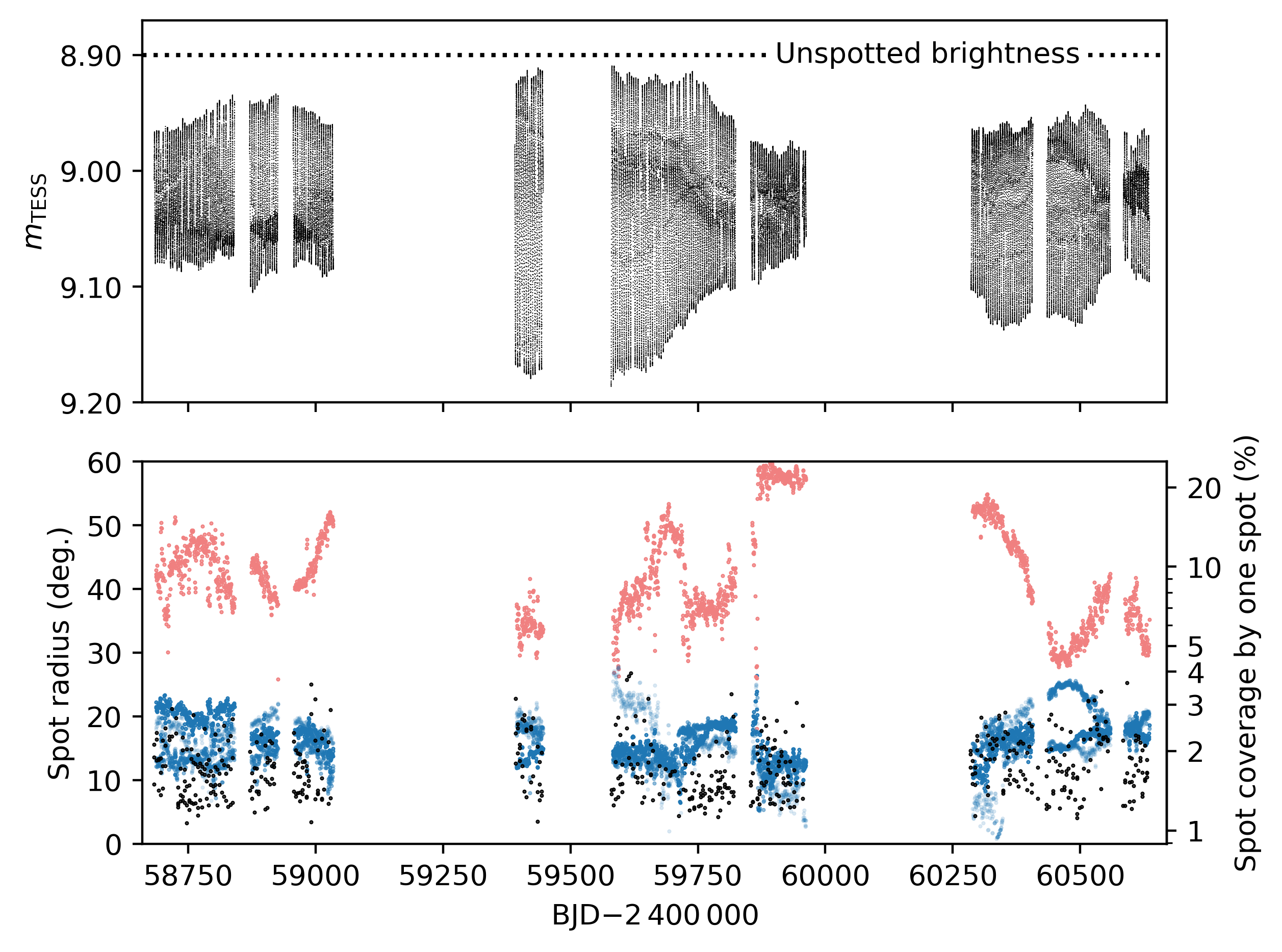}
      \caption{Top: Spot-only light curve with the unspotted brightness ($m_{\rm TESS} = 8.9$) marked as a dotted black horizontal line. Bottom: spot radii from time-series modelling: the three fixed equatorial spots are shown with blue dots with different shades while the free latitude spot is in red. The eclipse-mapping results are shown with larger black dots.}
      \label{fig:spot_coverage}
\end{figure}

The modelling was carried out in the reference frame of the binary orbit. Our spot model describes the spots with three parameters: latitude, longitude, and radius (the effective temperature of the spots is fixed).
In all of the nine light curve segments we used steps corresponding to 1.5~rotations of the primary. We tested different spot configurations, and four spots yielded the best balance between goodness of fit and model complexity. Three spots were fixed to $+2^\circ$ latitude based on the eclipse-mapping results, since there is very limited information on the spot latitudes from photometry alone. The fourth spot was allowed to move freely in latitude, most of the time covering the pole in different extents which accounts partly for the excess light variation above the rotational modulation as shown in the upper panel of Fig.\,\ref{fig:spot_coverage}. This spot  can explain some of the extra variation outside the eclipses, depending on its latitude, which is high most of the time. The free latitude and the constant spot temperature (see below) together increase the uncertainty of this spot's position and size, therefore, the fitted parameters of the fourth spot should be interpreted with caution.
Altogether we had nine free parameters: four spot longitudes, four radii, and one latitude. We used the following starting values for the spot parameters: $20^\circ$, $90^\circ$, $180^\circ$ and $270^\circ$ for longitude, $2^\circ$ for latitude, and $15^\circ$ for radius. The spot temperature was fixed to an average of 4100\,K originating from the eclipse modelling; the linear limb darkening coefficient was set to 0.55 \citep{2017A&A...600A..30C} taking into account the uncertainties of the stellar parameters given in Table\,\ref{tab:methods_comparison} (see Sect.\,\ref{sect:stellar_properties}).

\begin{figure}[t]
    \includegraphics[width=\columnwidth]{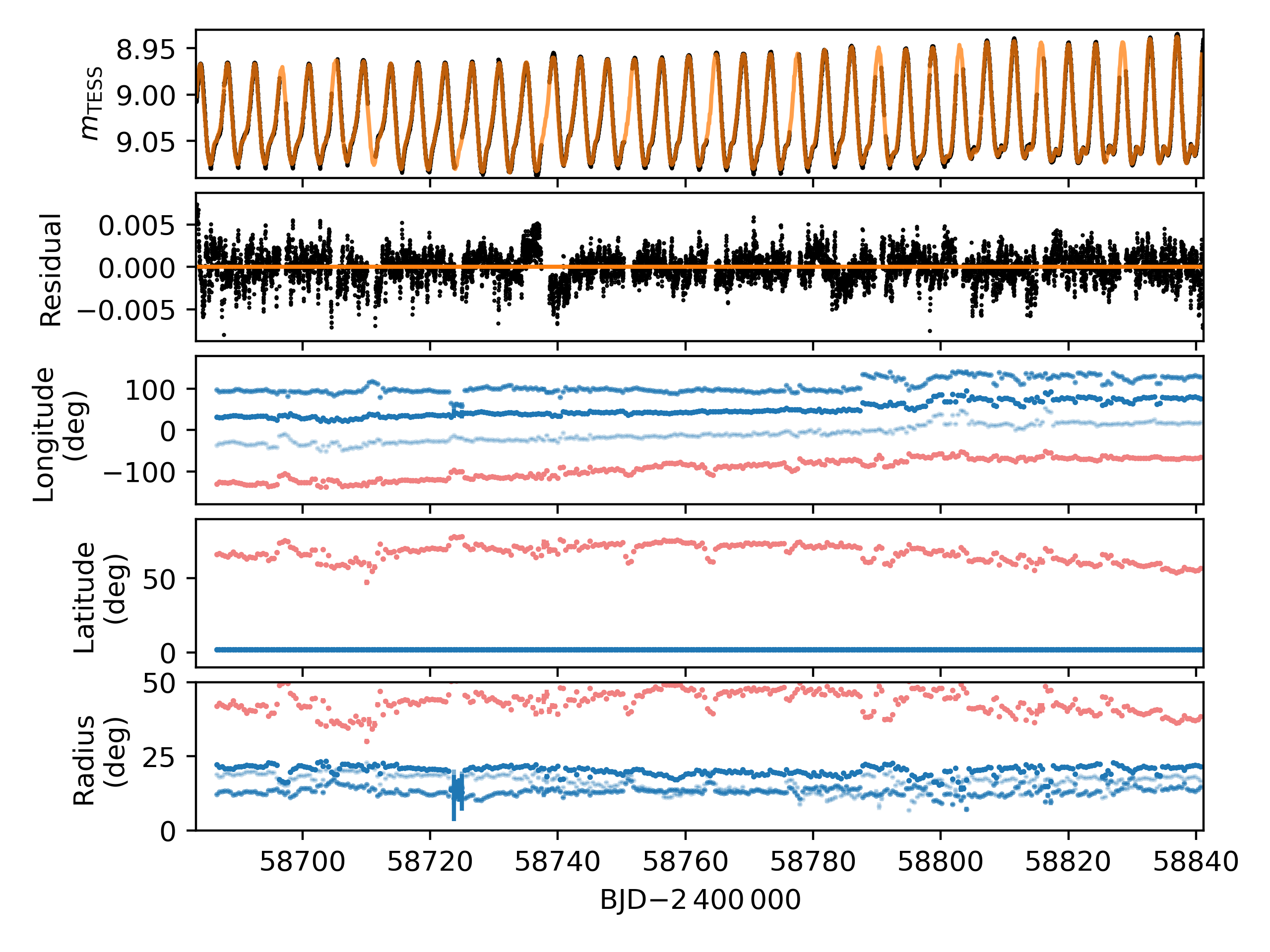}
      \caption{Time-series light curve fit for $\sim$160~days of the spotted light curve of V521\,Dra as an example. The first panel shows the data and the fit. The second panel shows the residuals after the model has been removed from the observations.
      In the three plots below, different colours show the resulting coordinates (longitude, latitude, radius) of the four spots, red colour marks the spot with free latitude solution. The small sudden deviations in the parameters correspond to gaps from missing data.}
      \label{fig:tserfit}
\end{figure}

The spot latitudes from full light curve modelling are shown together with the eclipse-mapping solutions, for comparison, on the right panel of Fig.~\ref{fig:tr_map}.
In Fig.\,\ref{fig:tserfit}, we show the full light curve model solutions from a $\sim$160\,d long segment of TESS observations.
All results from full light curve modelling are available in the Digital attachment.

Assuming more than one free latitude spot results in unstable latitude values which, in turn, effect the spot radii to some extent, whereas the longitudes remain stable. Experiments and details on the number of the applied spots in fitting light curves and in the stability of fitted latitude values with our method are given in \citetalias{2025olah} (Appendix A). 

\section{Stellar properties \label{sect:stellar_properties}}

\subsection{Radial velocity analysis \label{sec:rv_measurements}}

We performed radial velocity (RV) measurements on the spectra obtained with FIES, using ensemble absorption-line analysis techniques, which are generally reliable for slowly rotating stars. However, component A exhibits rapid rotation and strong variability in its line profile, due to the spots also detectable in the light curve, limiting the achievable RV precision despite high signal-to-noise ratios. These limitations reflect the systematic uncertainties introduced by the line profile variability, rather than limitations in data quality.

\begin{figure}[t!]
    \includegraphics[width=\columnwidth]{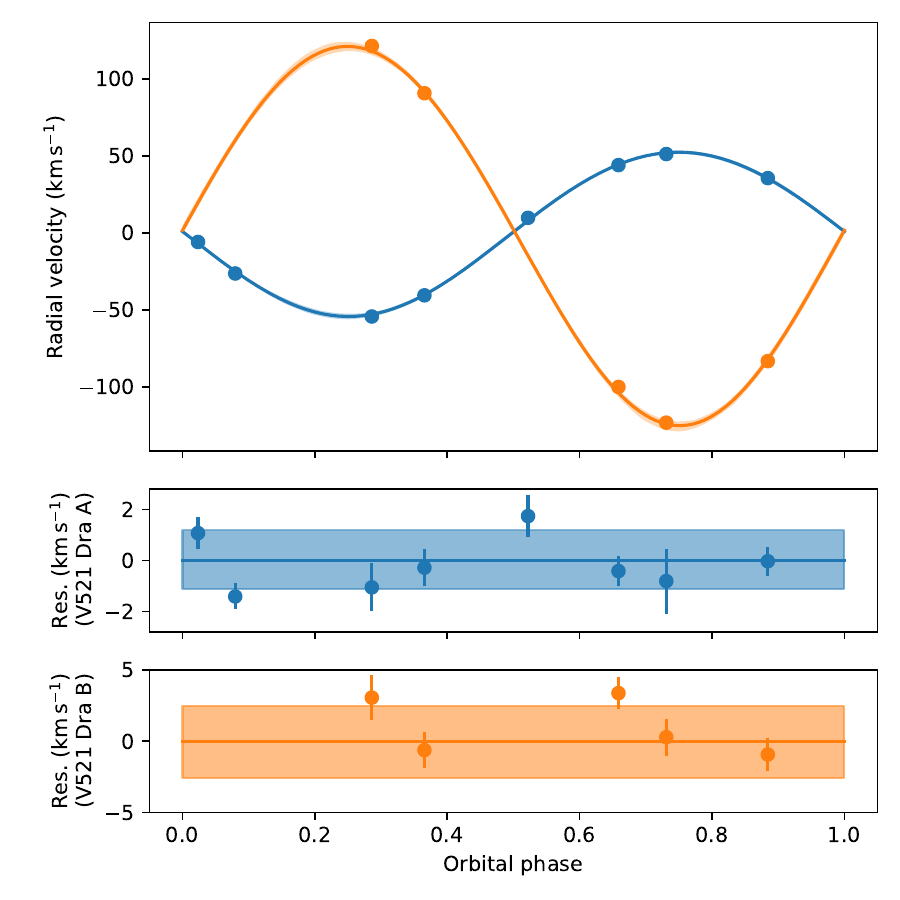}
      \caption{Top: Radial velocity phase curves of V521\,Dra~A and B with the best-fit Keplerian models in blue (component~A) and orange (component~B). The reference velocity of the system has been subtracted from the observations. Middle and bottom: Residuals after subtracting the best-fit Keplerian models from the primary and secondary, respectively. Shaded areas correspond to one sigma uncertainties in the model solutions.
      }
      \label{fig:phase_RV}
\end{figure}

The extraction of radial velocities from FIES spectra and the analysis of RV data are described in detail in Appendix\,\ref{app:rv_analysis}. Table\,\ref{table:radvelAB}  lists the barycentric Julian dates and radial velocities of the FIES observations. The RV measurements with the model posteriors are shown in Fig.\,\ref{fig:phase_RV}. We report the results in the fourth panel of Table\,\ref{table:1}.

We combined the results from the RV analysis with those from eclipse modelling. Using the component masses and orbital period, we calculated the semi-major axis $a$ of the system, which we then combined with the $a/R_A$ and $R_A/R_B$ parameters derived from eclipse modelling. The resulting parameters, along with those obtained from other analyses, are summarised in Table\,\ref{tab:methods_comparison}.

We estimated the chromospheric activity of the system from the cores of the Ca~II H\&K lines, obtaining a fractional emission flux of $\log R'_{\rm HK} = -3.9$. Appendix\,\ref{app:caHK} describes the chromospheric emission calculations in detail. Figure\,\ref{fig:caHK} in Appendix\,\ref{app:caHK} shows the spectra around Ca~II H\&K line from the FIES observations.

\begin{table*}[t!]
\centering
\caption{Binary system parameters derived from different analysis methods. \label{tab:methods_comparison}}
\begin{tabular}{lcccc}
\hline \hline
Parameter & Eclipse LC + RV & \textsc{Phoebe} & SED (no spots) & SED (spots) \\
\hline
\multicolumn{5}{l}{Binary star orbit} \\  \hline 
Orbital period $P_{\rm orb}$ (d) & 4.270337 & 4.270337 & 4.270337 & 4.270337 \\
Semi-major axis $a$ ($R_\odot$)
    & $14.9 \pm 0.3$
    & ---
    & $14.8 \pm 0.1$
    & $14.9 \pm 0.1$ \\
Inclination $i$ (deg)
    & $89.6^{+0.3}_{-0.4}$
    & $88.6 \pm 0.3$
    & ---
    & --- \\
Eccentricity $e$
    & $0.02 \pm 0.01$
    & $0.017 \pm 0.004$
    & ---
    & --- \\
Age (Gyr)
    & ---
    & ---
    & $1.6 \pm 0.1$
    & $1.6 \pm 0.1$ \\
\hline
\multicolumn{5}{l}{Star A} \\  \hline 
Mass $M_{\rm A}$ ($M_\odot$)
    & $1.7 \pm 0.1$
    & $1.7 \pm 0.1$
    & $1.68 \pm 0.05$
    & $1.69 \pm 0.05$ \\
Radius $R_{\rm A}$ ($R_\odot$)
    & $6.2 \pm 0.1$
    & $6.1 \pm 0.1$
    & $6.0 \pm 0.1$
    & $6.1 \pm 0.2$ \\
Effective temperature $T_{\rm eff,A}$ (K)
    & ---
    & 5011 (fixed)
    & $5010 \pm 50$
    & $5010 \pm 20$\tablefootmark{a} \\
Luminosity $L_{\rm A}$ ($L_\odot$)
    & ---\tablefootmark{b}
    & $21 \pm 3$
    & $20.6 \pm 0.3$
    & $19 \pm 1$\tablefootmark{c} \\
Surface gravity $\log g_{\rm A}$ (cgs)
    & ---
    & $3.09 \pm 0.05$
    & $3.10 \pm 0.02$
    & $3.10 \pm 0.02$ \\
\hline
\multicolumn{5}{l}{Star B} \\  \hline 
Mass $M_{\rm B}$ ($M_\odot$)
    & $0.74 \pm 0.04$
    & $0.74 \pm 0.05$
    & $0.73 \pm 0.03$
    & $0.74 \pm 0.03$ \\
Radius $R_{\rm B}$ ($R_\odot$)
    & $0.85 \pm 0.02$
    & $0.75 \pm 0.02$
    & $0.67 \pm 0.02$
    & $0.67 \pm 0.02$ \\
Effective temperature $T_{\rm eff,B}$ (K)
    & ---
    & $4880 \pm 80$
    & $4830 \pm 130$
    & $4850 \pm 140$ \\
Luminosity $L_{\rm B}$ ($L_\odot$)
    & ---\tablefootmark{b}
    & $0.29 \pm 0.04$
    & $0.23 \pm 0.04$
    & $0.24 \pm 0.04$ \\
Surface gravity $\log g_{\rm B}$ (cgs)
    & ---
    & $4.56 \pm 0.02$
    & $4.65 \pm 0.01$
    & $4.64 \pm 0.01$ \\
\hline
\end{tabular}
\tablefoot{
\tablefoottext{a}{Spot filling factor $f_s = 0.20 \pm 0.04$, temperature ratio of spotted and unspotted photosphere: $T_{\rm spot}/T_{\rm eff} = 0.86 \pm 0.03$.}
\tablefoottext{b}{Flux ratio in the TESS band $F_A/F_B = 55.8 \pm 0.4$.}
\tablefoottext{c}{Luminosities of the unspotted and unspotted surfaces: $16.7 \pm 1.2$ and $2.3 \pm 0.6~L_\odot$, respectively.}
}
\end{table*}

\subsection{Physical scale of the system and the components \label{sec:phoebe}}

The light curve of V521\,Dra was modelled using the binary star modelling code \textsc{Phoebe} \citep{prsa2005}, which is based on the Wilson--Devinney framework and provides a physically consistent treatment of Roche geometry, limb darkening, reflection effects, and gravity darkening. The solution was obtained by fitting the phase folded, spot-corrected light curve reconstructed from the Fourier-cleaned data (Sect.\,\ref{sec:disentangle}), adopting the mass ratio derived from the radial velocity analysis. The temperature scale of the primary component was anchored to the value inferred from the SED modelling (Sect.\,\ref{sec:LC+SED}), while the secondary's temperature was adjusted during the fitting procedure. Limb darkening coefficients were initially taken from tabulated values appropriate for the stellar parameters and were subsequently allowed to vary within the solution.

Because the photometric solution constrains the system in normalised units, the stellar dimensions were first obtained as fractional radii ($R/a$). The absolute semi-major axis and stellar masses were then derived from the spectroscopic radial velocity solution, thereby fixing the physical scale of the system. In this way, the eclipse morphology determines the geometric configuration, while the radial velocity amplitudes set the absolute dimensions. Multiple radius definitions (pole, point, side, back) were examined to quantify the effects of Roche distortion on the inferred stellar parameters. The relative radii to the pole, point, side and back are $0.39$, $0.45$, $0.41$, and $0.43~R_A/a$, respectively.

\subsection{Spectral energy distribution analysis \label{sec:LC+SED}}

As discussed in Sect.\,\ref{sec:rv_measurements}, our spectroscopic radial velocity observations and analysis have yielded the masses (to within a factor of $\sin^3i$) with associated uncertainties of $\sim$5\%. Our analysis of the orbital light curve  (Sect.\,\ref{sec:transit_mapping}) provided information on the inclination angle $i$, the relative radii $a/R_{\rm A}$, $R_{\rm A}/R_{\rm B}$, and the flux ratio $k_f$, in the TESS band, along with their uncertainties. The actual physical radii can be found by computing the orbital semimajor axis from the masses and period, and using that to convert the relative radii into absolute physical quantities. Moreover, even the age of the system can be estimated from the mass and radius of the primary star \citep[see, e.g.][]{yakut2025}. A convenient way to propagate the uncertainties in these estimates would be via Monte Carlo draws of the parameters and their uncertainties \citep[see, e.g.][]{joss_rappaport84}. The combined light curve and spectral energy distribution (SED) analysis is detailed in Appendix\,\ref{app:sed}.

\begin{figure}[t]
    \includegraphics[width=\columnwidth]{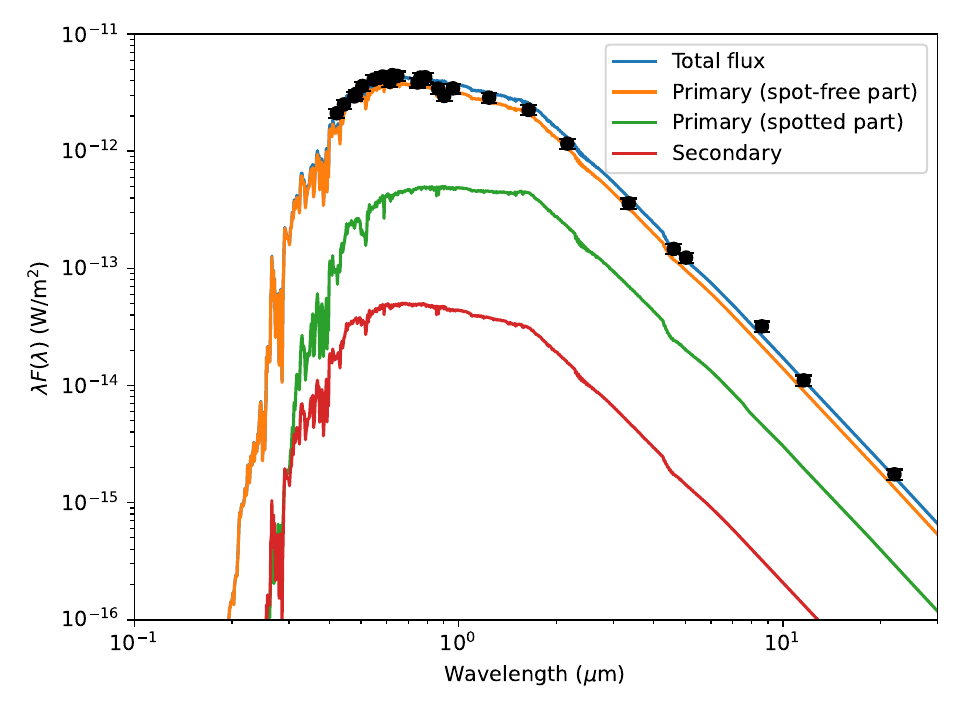}
      \caption{SED data of V521\,Dra (black dots) along with the best-fit solution, using a three component model. The components correspond to the spotless (orange) and the spotted (green) parts of the primary's surface and the secondary (red). The blue curve shows the sum of stellar contributions.}
      \label{fig:sed}
\end{figure}

Two sets of sampling were run on the SED of the system, one in which the primary was treated assuming that there are no spots on its surface, and one in which a fraction of the primary was covered by spots. In this second model, a fraction of the total surface area of the primary $f_s$ (spot filling factor) is occupied by spots, with the spots having a mean temperature ratio of $t_r = T_{\rm spot}/T_{\rm star}$. We sampled $f_s$ and $t_r$, using uniform priors, between $(0.1,0.25)$ and $(0.8,0.95)$, respectively. The prior ranges were motivated by the results of the full light curve modelling for $f_s$ and eclipse mapping for $t_r$. We obtained a spot filling factor of $f_s = 0.20 \pm 0.04$, and a spot temperature ratio of $t_r = 0.86 \pm 0.03$. The results from both models are listed in the last two columns of Table\,\ref{tab:methods_comparison}. The best-fit solution to a three component SED model (spot-free part of the primary, spotted part of the primary, and the secondary) is shown in Fig.\,\ref{fig:sed}.

Finally, we note that we carried out fits for two cases in regard to the distance to the source. In one case we restricted the priors on the distance to $\pm 3 \sigma$ of the cited Gaia distance uncertainty \citep[$322 \pm 4$\,pc,][]{bailer-jonesetal21},. In the other case we opened up the prior so as to independently determine the distance,  obtaining the estimate $324 \pm 11$ pc (1-$\sigma$), in agreement with Gaia's distance, but much less accurate.

\subsection{Temporal change (O-C) of the orbital period}

\begin{figure}[t!]
    \includegraphics[width=\columnwidth]{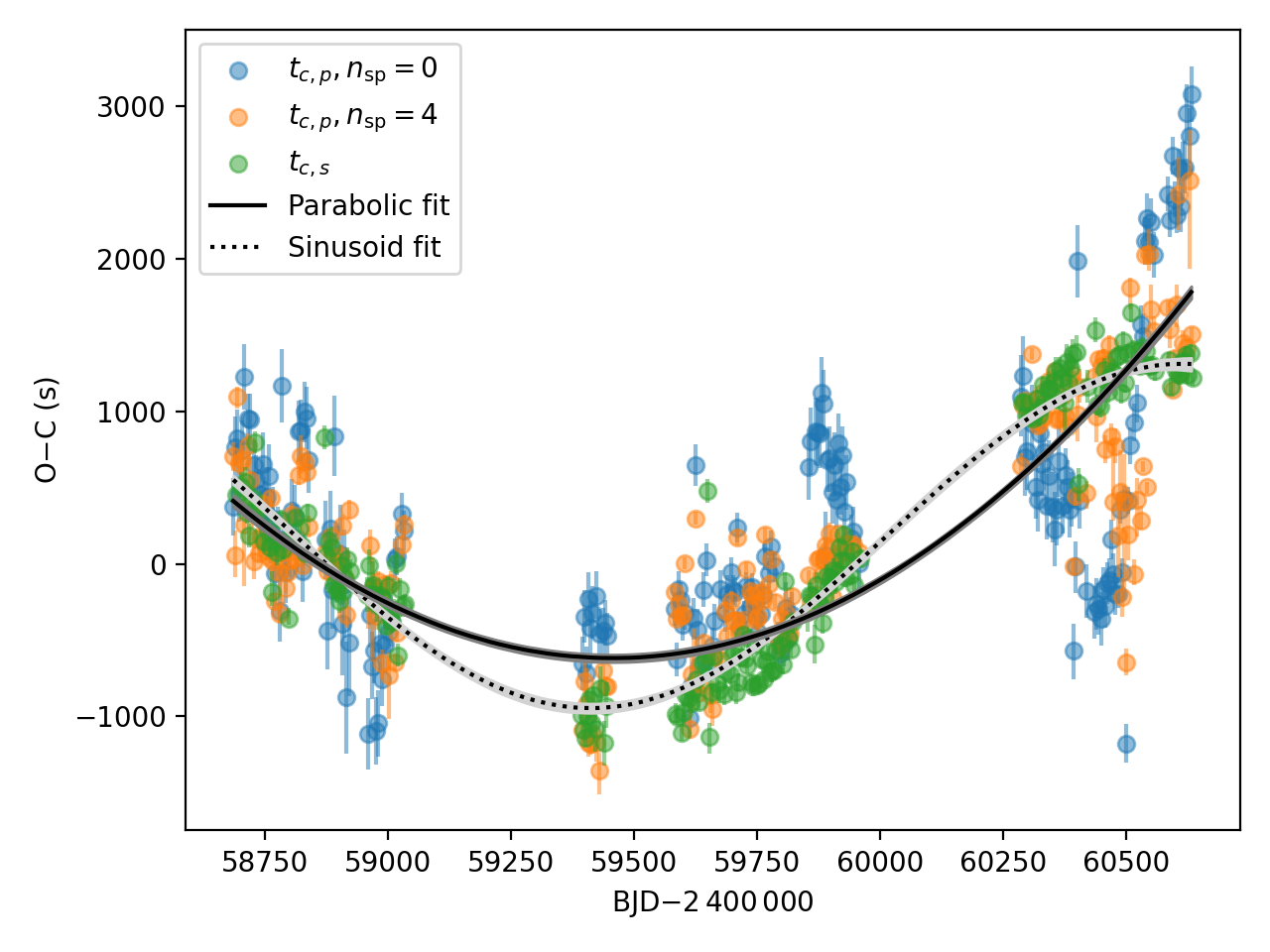}
      \caption{O--C curve of the eclipse mid-times. Primary eclipses are shown with blue and orange dots, based on the numbers of spots, $n_{\rm sp}$ assumed during eclipse modelling (blue for $n_{\rm sp}=0$ and orange $n_{\rm sp} = 4$). Secondary eclipses are marked green. A parabolic and a sinusoidal model, corresponding to a secular change in the orbital period and the presence of a third body on a wider orbit, are shown with continuous and dashed black lines.}
      \label{fig:oc}
\end{figure}

Starspots have long been recognised as a source of systematic errors in eclipse mid-time measurements, introducing jitter in the observed-minus-calculated (O--C) curves of binary systems \citep[see, e.g.][]{kalimeris2002}. We analysed the O--C diagram constructed from the eclipse mid-times of the system and found that the measured mid-times $t_c$ depend strongly on the number of spots, $n_{\rm sp}$, assumed in the model. Fig.\,\ref{fig:oc} shows the O--C curve, with each eclipse mid-time colour-coded by the number of spots used. All eclipse mid-times are provided in the Digital attachment. For the fixed-baseline sample, the mean difference between the minimum and maximum $t_c$ estimates per eclipse was $\sim$820\,s, with a maximum difference of $\sim$3050\,s. The largest discrepancies occured in the final segment of continuous observations (BJD $> 2\,460\,000$), when large, high-contrast spots cross the eclipse chord.

The O--C curve exhibits a clear temporal modulation, which could represent either a parabolic trend or the lower segment of a single sine wave. A paraboloid would indicate a secular increase in the orbital period, while a sinusoidal would imply dynamical influence of an outer, third-body. The best-fit paraboloid and sinusoidal solutions are shown in Fig.\,\ref{fig:oc}. Confirming the nature of the modulation remains challenging, as the primary light curves' mid-eclipse times suffer from highly correlated noise from starspots that our model can only partially account for. The final segment of the O--C curve is particularly problematic due to the presence of a large spot during ingress. If the modulation is sinusoidal, the absence of a complete cycle in the TESS data leaves its periodic nature unconfirmed.

We investigated if a third body can be present in the system, using the SED data, by introducing an extra star with fixed mass while maintaining the original constraints on the primary binary. Assuming coeval evolution, the mass and age of the third star were used to determine its radius and effective temperature. SED-fits were computed for third-star masses of $0.7, 0.9, 1.1, 1.3, 1.5$, and $1.6~M_\odot$. For masses up to $1.3~M_\odot$, the inferred properties of the binary remained essentially unchanged, though spectroscopy could rule out a hot third star contributing significantly to the total system light. Even at $1.5~M_\odot$, where the third star contributed $\sim$37\% of the system light and dominates the blue flux, the fitted semi-major axis to primary radius ratio remained unaffected. We conclude that a third star of up to $\sim 1.1-1.2~M_\odot$ could exist undetected. The Gaia Renormalized Unit Weight Error \citep[RUWE,][]{lindegren2018} parameter for the system is $1.07$, indicating no evidence for a third star in the system. The presence of a third body in the system, however, would not be completely unexpected. \citet{tokovinin2006} estimates the fraction of systems with a tertiary companion to be around $0.7$ for binaries with orbital periods $P \sim 5$~d.

The presence of a third light in the system would cause the eclipse light curves to look shallower than they actually are, meaning that the parameters $k_r$ and $k_f$ would be underestimated. For example, if a $1.3~M_\odot$ third body, with a $15\%$ contribution to the flux of the whole system would cause the $k_f$ and $k_r$ parameters to be underestimated by $15\%$ and $7\%$, respectively. The spot properties derived from eclipse mapping would remain unaffected if a third light was present in the system. More light will be shed on the course of the O--C curve when TESS will resume observing the northern ecliptic hemisphere in its ninth year of operation, starting in September~2026.

\section{Discussion \label{sect:discussion}}

\subsection{Rotation of the primary
\label{sect:discussion_rotation}}

We found a subtle but significant difference between the orbital and the average rotational period of the primary. This feature could be related to the differential rotation of the primary star, or alternatively, the system is not fully synchronous.

Assuming that the measured average rotation period from all data $\overline{P}_{\rm rot}=4.257$\,d corresponds to the equatorial rotation period and that the star exhibits solar-like differential rotation (that is, the equator rotates the fastest and the angular velocity decreases continuously towards the pole), a synchronised surface latitude matching the orbital period $P_{\rm orb}=4.27$~d can exist only if the relative surface shear is at least $\lvert\overline{P}_{\rm rot}-P_{\rm orb}\rvert/\overline{P}_{\rm rot}=0.003$. However, in this case the synchronisation would occur only at the pole (i.e.~the polar rotation period equals the orbital period), which is unlikely, so a somewhat larger surface shear is more plausible in order to produce a synchronised belt at mid latitudes. For example,  shear parameters of 0.005 and 0.01 would produce a co-rotating belt at typical mid latitudes of $\sim$50$^{\circ}$ and $\sim$30$^{\circ}$, respectively.

The difference between the rotational and orbital periods may indicate a so-called pseudo-synchronous rotation, which means close to synchronism between the rotation and revolution near periastron \citep{1981A&A....99..126H}. From the known values of the orbital and rotational periods a small eccentricity value of $e \sim 0.03$ follows. This value is consistent with our present result of $e \sim  0.02 \pm 0.01$ from modelling light curve+radial velocity data (cf. Table\,\ref{tab:methods_comparison}). 

An alternative explanation may be an azimuthal dynamo wave. This would manifest itself as an apparent drift of active regions, with respect to the stellar rotation. Such waves have been reported from both numerical models \citep{cole2014} and observational studies \citep{hackman2012}. In summary, possibly all scenarios, i.e.~non-complete synchronisation, differential rotation and an azimuthal dynamo wave could contribute to the apparent drift of the spots in the orbital reference frame.

\subsection{Evolution of spots from eclipse mapping \label{sect:spot_evolution}}

We visually identified $12$ spot groups exhibiting recurring signatures that persisted for at least five eclipses and drifted at least $6.5^\circ$ in longitude in the orbital frame. These features enabled us to constrain the angular velocity of the drift of spots and provide insight into spot evolution over extended timescales. In addition to spot parameters from eclipse mapping, the flux deficit caused by individual spots as it co-rotates with the stellar surface can be estimated using Eq.~7 of \citet{tregloan2019}. Fig.\,\ref{fig:eclipsemap_coloured} in Appendix\,\ref{app:spotevolution} presents a colour-coded version of the eclipse-mapping time series, with different colours for each recurring spot group signal. The evolution of radii, contrasts, flux deficits and longitudes for individual spot groups is shown in Fig.\,\ref{fig:spotevolution_appendix}.

The spot group with the longest traceable signal in the eclipse-mapping data persisted for $256$~days, placing a lower limit  on the maximum lifetime of spot groups on V521\,Dra~A. While polar spots with lifetimes from months to years are commonly identified on RS\,CVn stars using Doppler and Zeeman--Doppler imaging \citep{strassmeier1990, lindborg2011}, they are less confidently identified closer to the equator. Latitude estimates for these lower-latitude features are less precise than their longitudes, due  to the ambiguity in determining on which hemisphere the spots are located when using Doppler imaging. Eclipse mapping constrains the scanned latitude band to $(-7.0_{-0.6}^{+1.0}\,^\circ, 8.9_{-0.6}^{+1.0}\,^\circ)$, which gives more accurate latitude information on the spots than most other methods. However, we stress that not even eclipse mapping can give precise spot latitudes, the only constraint we have is that spots are somewhere within this $\sim$$16^\circ$ wide latitude range. Therefore, we cannot track the evolution of spot latitudes for example.

In theory, it is possible to track the lifetimes of spot groups on even longer baselines, by estimating when a spot group should reappear in the eclipse band and matching the appearance of spots to that time. However, due to the uncertainties involved in estimating the angular velocities of spots and possible spot migration, we chose not to go into such uncertain regimes.

The spot signals from the eclipse mapping show clearly distinct angular velocities (see Fig.\,\ref{fig:tr_map}).  These features originate from a narrow latitude range near the equator . However, differential rotation laws are relatively flat near the equator, as they scale with the sine of the latitude. Therefore, latitudinal differential rotation is unlikely to account for the differing angular velocities. Instead, the variation in angular velocities is more plausibly explained by radial differential rotation within the convective envelope, with spots anchored at different depths. To quantify these differences, we fitted a first-order polynomial to the longitude drift of the spots using linear regression, in order to obtain their angular velocities in the orbital reference frame. We find angular velocities in the orbital reference frame between $(2.4 \pm 0.5) \times 10^{-3}$ and $(7.2 \pm 0.4) \times 10^{-3}$~rad\,~d$^{-1}$. Such angular velocity difference would give a relative shear of $k = \Delta \Omega/\Omega = 0.0033 \pm 0.0005$. This number represents a lower limit  for the shear, as we cannot know how deeply each active region is anchored in the convective layer of the star.  Radial shear measurements for these stars have practically been unavailable so far. \citet{jermyn2020} analysed the rotational and orbital periods of $816$ eclipsing binary systems observed by the Kepler space telescope and found that for short-period binaries there is very little surface shear.

\subsection{Spot coverage}\label{sec:spot_coverage}

A full quantitative comparison of eclipse mapping and full light curve modelling of spots is challenging due to their inherent differences. Eclipse mapping provides independent spot contrasts and radii but covers only a limited longitude and latitude range on the primary. Full light curve modelling of spots reconstructs the entire surface but has limited latitude accuracy and, with single-band photometry, suffers from degeneracy between spot radii and temperatures. However, inferred properties such as spot coverage are more directly comparable.

By summing  spot areas, we can estimate the  stellar spot coverage. In  full light curve modelling four spots were used, three fixed near the equator,  and one with free latitude to account for the overall light variability (see Sect.\,\ref{full_lc}). While the three equatorial spots have about $16^{\circ}$ radii on average, the fourth has a wider size range between $18^{\circ}-60^{\circ}$ radii, which also depends on the latitude. The three equatorial spots together cover $6 \pm 1\%$, while the fourth spot covers about $13 \pm 5\%$ of the surface (its higher uncertainty is explained in Sect.~\ref{full_lc}), meaning that the overall spot coverage is nearly $20\%$. 
The spot filling factor from spot modelling is consistent with the SED-fit results (see Sect.\,\ref{sec:LC+SED}). We note that the $\sim$20\% from spot modelling is a lower limit, since the true unspotted brightness of the star is not known. Taking into account that the secondary star makes up $L_B/(L_A + L_B) = 0.012$ from the whole luminosity of the system (see Sect.\,\ref{sec:LC+SED}), the resulting spot radii would be larger by about $0.1^\circ$, and the overall spot coverage increases only by a negligible amount.

Eclipse mapping gives an average spot radius of $11.2^{\circ}$, corresponding to $\sim$1\% spot coverage from a single spot within the eclipse band. Individual eclipse maps with one to three spots reach a spot coverage of a few per cent. Full light curve modelling gives an average spot coverage of about $4\%$ in the same region (Fig.\,\ref{fig:tr_map}, right panel), in good agreement with the eclipse-mapping results. The remaining differences may partly reflect that full light curve modelling assumes a fixed spot temperature, whereas eclipse mapping treats spot contrast as a free parameter.

Figure\,\ref{fig:spot_coverage}  shows the spot-only brightness and the corresponding spot radii and coverage from both the full light curve modelling and eclipse mapping. The spot-only full light curve modelling is free from eclipses (see Fig.\,\ref{fig:example}), so an overall similarity of the spot coverage and spot radii from the two approaches around the eclipse area of the primary is promising.

The agreement between the spot longitudes and their drifts inferred from eclipse mapping and full light curve modelling also implies the consistency of the two methods (Fig.\,\ref{fig:tr_map}, right panel). The longitudes from full light curve modelling either coincide with the highest contrast spots from eclipse mapping or lie between two eclipse-mapping spot longitudes. In the latter case, the lower spatial resolution of the full light curve modelling likely prevented it from resolving two nearby spots that are recovered separately by eclipse mapping.

\subsection{Evolutionary state of the system}

\begin{figure}[t!]
\centering
\includegraphics[width=\columnwidth]{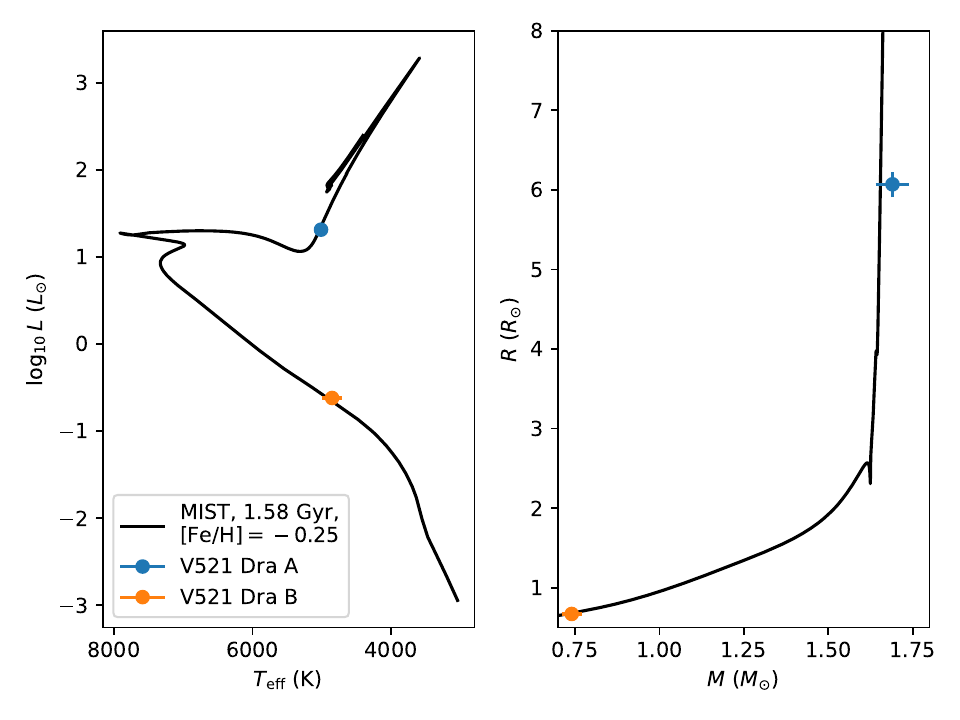}
\caption{Comparison of the stellar properties of the binary and a MIST isochrone with [Fe/H]$=-0.25$, age $=1.58$~Gyr, and initial rotation $v/v_{\rm crit}=0.4$. Left: Hertzsprung--Russell diagram. Right: mass-radius diagram.}\label{fig:mr_hrd}
\end{figure}

We compared the measured properties of the binary components with theoretical MIST isochrones\footnote{\url{https://waps.cfa.harvard.edu/MIST/}}. The components are shown in a mass--radius and Hertzsprung--Russell diagram together with the corresponding isochrone in Fig.\,\ref{fig:mr_hrd}. From the properties of the primary (the red giant), it is clear that it resides on the hydrogen-shell–burning RGB. Consequently, the precise age of the stellar system derived in Sect.\,\ref{sec:LC+SED} effectively corresponds to the total main-sequence lifetime of the  red giant progenitor. This quantity depends primarily on the stellar mass and metallicity, which is also reflected in the nearly vertical behaviour of the isochrone in the mass--radius diagram.

This age estimate relies on the assumption that the current mass of the RG is equivalent to the initial mass. Based on the \textsc{Phoebe} modelling, the point of the primary (the side facing the secondary) essentially fills its Roche lobe ($r_{\rm point} = 0.45~R_A/a$, $r_{\rm Roche} \sim 0.45$), and this assumption is therefore fundamentally uncertain. However, it is encouraging that the measurements of both the RG and the companion main-sequence star are in agreement with the same isochrone.

Due to the ascent of the RG along the RGB, the binary is likely preceding a short-lived evolutionary phase of mass transfer and possibly common-envelope evolution. Systems such as this are intrinsically rare to observe, due to the rapid stellar expansion on the RGB. A recent population-synthesis study by \citet{mazzi2025} found that four fifths of all RG binary interaction products are in the core-helium burning or asymptotic giant branch phases, rather than on the RGB, implying that binary disruption of these systems typically occurs during the RGB ascent. To our knowledge V521\,Dra is the only system observationally confirmed to be at this critical point of no return. This is a crucial stage to model accurately in binary evolution simulations, and V521\,Dra can therefore offer unique constraints on the behaviour of such binaries immediately before disruption.

\begin{figure}[t]
    \includegraphics[width=\columnwidth]{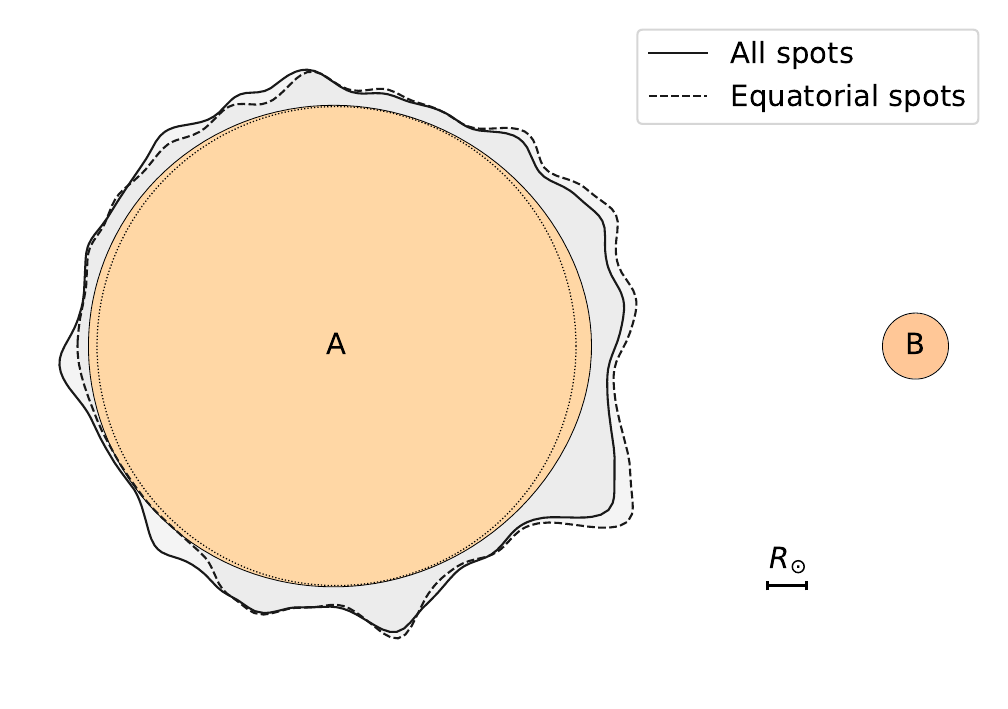}
      \caption{Distribution of the spot longitudes from the full light curve modelling, visualised by a kernel density estimate wrapped around the projected shape of the primary star. The binary system is shown to scale, also with realistic stellar colours from \citet{2021AN....342..578H}. An overabundance of spots is visible on the side facing the secondary star.}
      \label{fig:primary_shape}
\end{figure}

The shape of the primary star is slightly distorted due to the proximity effects as  detailed in Sect.\,\ref{sec:phoebe}. The shape of the primary is illustrated in Fig.\,\ref{fig:primary_shape}, calculated from the \textsc{phoebe} forward model. To estimate the distortion a triaxial ellipsoid was fitted to the \textsc{phoebe} model mesh, yielding relative axes of 1.068, 0.938, and 0.994. Figure\,\ref{fig:primary_shape} also shows the distribution of spot longitudes from the full light curve modelling, using Gaussian kernel density estimation. The spots show signs of clustering on the side facing the secondary, especially in the case of the three spots fixed near the equator during modelling. The circular mean of all spot longitudes from the full light curve modelling (solid line in Fig.\,\ref{fig:primary_shape}) is $-11^\circ \pm 3^\circ$, with a formal uncertainty from bootstrap resampling. This value is close to the nearest point to the secondary in the orbital reference frame (i.e.~zero longitude).  This suggests a connection between the spot positions and the shape of the star, as the magnetic flux tubes erupt in the common gravitational field of the system. More background of this phenomenon in different types of binaries, and some observational results are found, for example, in \citet[][and references therein]{2006Ap&SS.304..145O, 2007IAUS..240..442O}, and in \citet{2024MNRAS.529.4442S}.

\section{Conclusion}

This work is the second instalment in our series on mapping starspots on giant stars in eclipsing binaries, using eclipse mapping combined with full light curve modelling. Since \citetalias{2025olah}, we have advanced our methodology by isolating the orbital light curve from stellar activity, enabling a clearer characterisation of spot signals across the full light curve. This was necessary due to the significant binarity effects in the system's light curve. We have also implemented a rigorous set of inversion methods to thoroughly characterise the stellar system.

In this study, we analysed TESS observations of V521\,Dra spanning over 5.3~years, with an emphasis on starspot signatures. Our findings reveal several notable phenomena:

-- Starspot signatures detected in the eclipse band persist over several eclipses, exhibiting a continuous longitudinal drift. We have tracked some starspots for more than 250~days (60~stellar rotations). The lifetimes of these spots exceed significantly those observed on the Sun. 

-- We identified a subtle discrepancy between the orbital period and the rotational period of the spots from full light curve modelling. Assuming spin--orbit coupling, this discrepancy suggests differential rotation, or the presence of an azimuthal dynamo wave, or both.

-- The eclipse-mapping data indicates differences in angular velocities among co-rotating spots. Given that eclipse mapping is limited to a narrow latitude range around the equator ($\pm 10^\circ$), surface differential rotation is unlikely to account for these differences; instead, shear within the convective layer presents a more plausible explanation. 

-- The eclipse mid-times exhibit a non-linear modulation over the TESS observation period, persisting above the correlated noise generated by the spots in the eclipse band. The physical mechanisms underlying these variations remain unclear and require further investigation.

To extend this research, we will analyse a set of 25 additional eclipsing binaries from \citetalias{2025olah}, although with fewer eclipse light curves available for each system. By expanding our sample to a larger number of objects, we aim to uncover the underlying trends that govern the magnetic activity of this rare and intriguing class of stellar systems. Future analysis will provide deeper insights into the mechanisms driving starspot behaviour and their broader implications for stellar magnetism.

\section*{Data availability}

The spot parameters from time-series spot modelling and eclipse mapping, and the mid-eclipse times are available in electronic form at the CDS via \url{http://cdsweb.u-strasbg.fr/cgi-bin/qcat?J/A+A/...}.

\begin{acknowledgements}
Based on observations made with the Nordic Optical Telescope, owned in collaboration by the University of Turku and Aarhus University, and operated jointly by Aarhus University, the University of Turku and the University of Oslo, representing Denmark, Finland and Norway, the University of Iceland and Stockholm University at the Observatorio del Roque de los Muchachos, La Palma, Spain, of the Instituto de Astrofisica de Canarias.
We thank the anonymous reviewer for the helpful suggestions that improved the content and readability of the paper. AH~and~MT~acknowledge support from the Jenny and Antti Wihuri Foundation.
This work received funding from the Research Council of Finland (project PEAS, decision No.~376613).
The authors wish to acknowledge CSC – IT Center for Science, Finland, for computational resources.
This work was supported by the Hungarian National Research, Development and Innovation Office \'Elvonal grant KKP-143986. JST acknowledges support from Bologna University, "MUR FARE Grant
Duets CUP J33C21000410001". JHT~acknowledges Horizon Europe ERC grant no.~101125877.
This work made extensive use of \texttt{numpy} \citep{numpy}, \texttt{scipy} \citep{scipy}, \texttt{pandas} \citep{mckinney-proc-scipy-2010} and \texttt{matplotlib} \citep{matplotlib}.

\end{acknowledgements}

\bibliographystyle{aa}
\bibliography{biblio.bib}

\begin{appendix}

\section{Eclipse light curves \label{app:eclipse_lcs_method}}

To determine the orbital period $P_{\rm orb}$ and the epoch $T_0$ (defined as the mid-time of the first primary eclipse in the TESS time series), we used the same iterative method as in \citetalias{2025olah}. This method relies on first estimating the orbital period of the system using the Box-fitting Least Squares (BLS) algorithm \citep{kovacs2002}, and then identifying the eclipse mid-times in the TESS time series and sampling them with a linear regression with Markov Chain Monte Carlo (MCMC). We only used the secondary eclipses at this stage, as spot occultations during primary eclipses  added an extra level of uncertainty on the eclipse mid-times. 

To analyse the flux variations from spot occultations during primary eclipses, longer photometric variations on timescales from half to a full orbital period need to be removed from around the eclipse. In \citetalias{2025olah}, this was done by extracting segments from the light curve, centred around the eclipse mid-points, with a window size of $3\,T_{14}$ where $T_{14}$ is the eclipse length, leaving one eclipse length's worth of brightness measurements both before and after the eclipses. The out-of-eclipse variations, primarily from the rotation of the primary component, were then removed by a low order polynomial.

This method, however, only gives consistent results when the out-of-eclipse flux variations are relatively simple and the eclipses are short compared to the orbital period ($T_{14}/P_{\rm orb} \lesssim 0.1$). Since for V521\,Dra $T_{14}/P_{\rm orb} = 0.15$, and the out-of-eclipse variations are relatively complex due to the interplay of binarity effects and starspots, we reduced the window size to $1.3\,T_{14}$ and normalised the out-of-eclipse parts with a third order polynomial. This choice of parameters ensures consistent eclipse depths, while including as many out-of-eclipse datapoints as possible.

\section{Eclipse baseline and sampling \label{app:eclipse_baseline}}

We fixed the eclipse baseline for a number of practical reasons. First, the normalisation routine applied to the eclipse light curves works well for the primary eclipses, but has limitations for the secondary ones. In some secondary eclipses we find modulations in flux during the constant brightness phase, implying imperfect normalisation. Attempting to fit these features would give unrealistic eclipse baselines and would complicate the identification of spot signals. In addition, some primary light curves contained such long and high amplitude spot modulations that they obscured most of the eclipse baseline. In extreme cases, this can lead to the spots being mistaken for the baseline eclipse profile itself. Finally, by fixing the unspotted eclipse baseline, we restrict the sampling to the primary eclipses, which lowered computational costs by a factor of two.

To obtain the eclipse baseline, we sampled five primary and five secondary eclipses from different TESS sectors. We selected such primary eclipses whose central parts seem to show low spot modulations. This was to ensure that the fitted baseline is close to the physical reality. We used a model with two spots per primary eclipse. We fixed the orbital period to the value from the BLS analysis, and assumed non-eccentric orbits. We sampled the posterior probability densities with the Adaptive Metropolis (AM) algorithm of \citet{haario2001}. We sampled the light curves simultaneously using a first order moving average noise model (MA(1)) with independent noise parameters for each TESS sector to filter possible correlated noise in the observations. Parameter estimates are reported in the third panel of Table\,\ref{table:1}. We fixed the eclipse baseline to the mean of the posterior.

Similarly to the eclipse baseline, we sampled the spot solutions for eclipse mapping with the AM algorithm, using an MA(1) noise model, with independent noise parameters for each TESS sector.
Spot properties were estimated using the equations of \citet{silvavalio2008} and \citet{silvavalio2010}, see \citetalias{2025olah}. The spot radii in the eclipse band can be calculated in two different ways: based on half of the great-circle distance of the stellar surface points covered by the centre of the secondary at times $t_{\rm in}$ and $t_{\rm out}$, and between those covered at times $t_{\rm in} - \sigma_{\rm spot}$ and $t_{\rm out} + \sigma_{\rm spot}$ \citep[see][]{haris2025}. We refer to this second definition as the spot radius unless otherwise stated. The spot contrast is defined as $c = 1 - I_{\rm spot}/I_{\rm star}$, where $I_{\rm star}$ and $I_{\rm spot}$ are the intensities of the unspotted and spotted photosphere, respectively.

Given the stellar effective temperature, the weighted mean effective temperature of the occulted dark region can be estimated under the blackbody approximation \citep{silvavalio2010}:
\begin{equation}\label{eq:spot_temp}
T_{\rm s} = \frac{h \nu}{K_{\rm B}} \left \{ \log \left [ 1 + \frac{\exp \left ( \frac{h \nu}{K_{\rm B} T_{\rm eff}} \right ) - 1}{f} \right ]\right \}^{-1}\,,
\end{equation}
where $K_{\rm B}$ and $h$ are the Boltzmann and Planck constants, respectively, $\nu$ is the frequency corresponding to the central wavelength of the observing instrument (786.5~nm for TESS), $f = 1-c$ is the spot intensity relative to the central stellar intensity, and $T_{\rm eff}$ is the stellar effective temperature.

\begin{figure*}[hb]
\begin{center}
   \includegraphics[width=0.9\textwidth]{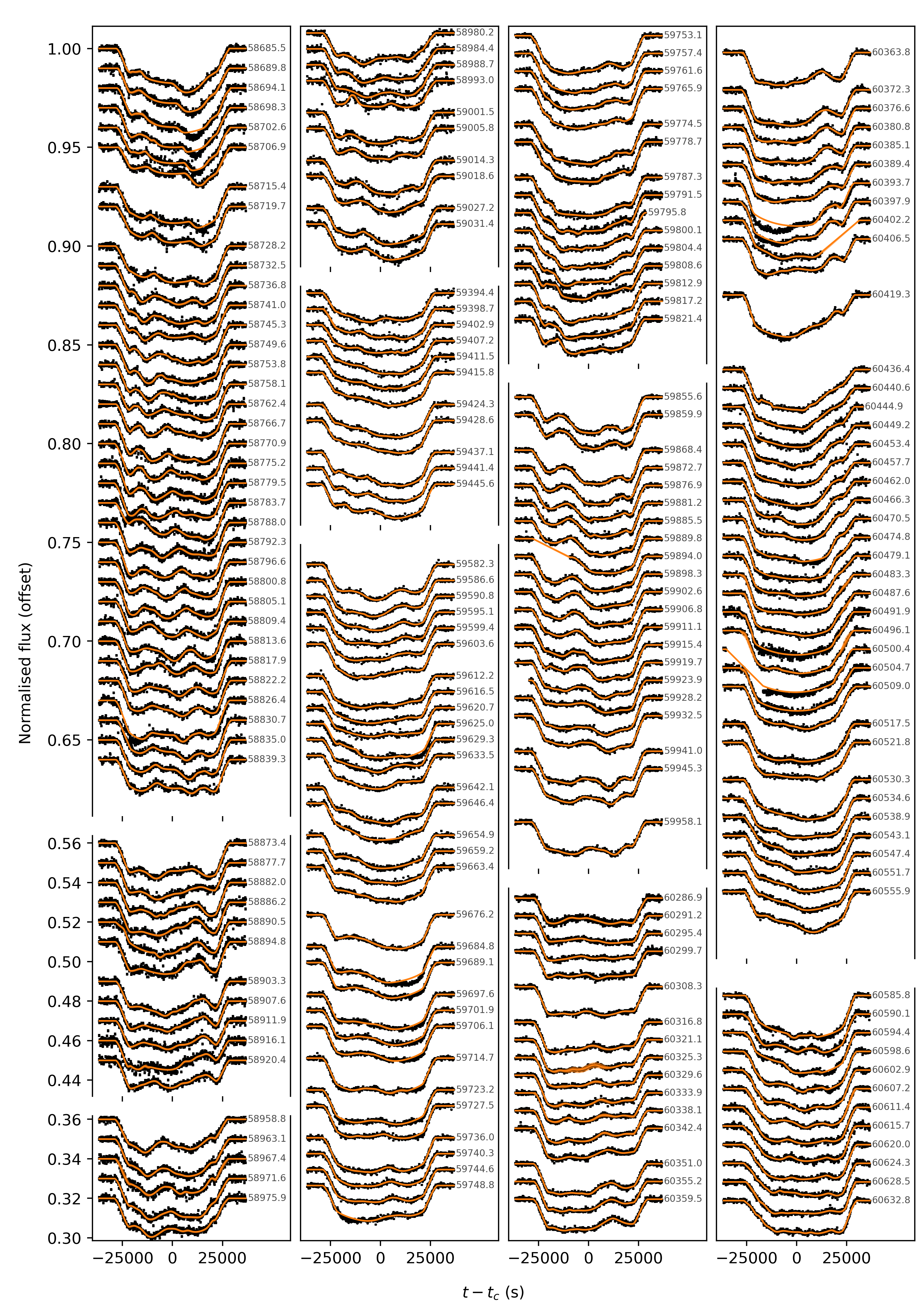}
      \caption{All analysed primary eclipse light curves of V521\,Dra. Best-fit eclipse models are shown in orange. The mid-eclipse times $t_c$ of individual eclipses are shown in terms of BJD$-2\,400\,000$ on the right of each eclipse. The figure is split into segments  when at least $5$ consecutive primary eclipses are missing from the analysis.}
      \label{fig:all_eclipses}
\end{center}
\end{figure*}

In Fig.\,\ref{fig:all_eclipses}, we show all eclipse light curves analysed with eclipse mapping.

\onecolumn

\clearpage
\twocolumn

\section{Separation of orbital modulation and stellar activity \label{app:separating_lc}}

The light curve of any eclipsing binary is composed of (i)~permanent periodic features such as eclipses, ellipsoidal light variations, illumination effects, and Doppler boosting, as well as (ii)~stochastic variations due to stellar activity (including spots) that may occur on many different timescales, including ones near the orbital period. To separate these components, we utilised the extensive TESS dataset, which spans nearly 2000 days, with observations covering roughly half of that period. This long baseline provides an opportunity to robustly disentangle the periodic orbital signals from stellar activity. Our approach followed the now well-established method used by researchers to isolate stellar pulsations from orbital features in binary systems.

First, we determined the long-term best orbital period from a coherent BLS transform of the entire TESS data set. In this case the period we found is $P_{\rm orb} = 4.27036 \pm 0.00004$\,days. We then chose the number of orbital harmonics, $N$, to utilise. The more harmonics, the better one can reproduce the short-timescale features such as the eclipse ingress and egress. On the other hand, the more harmonics one uses, the more likely spurious high-frequency statistical noise enters the light curve. For V521\,Dra we chose $N = 40$, which corresponds to accurately reproducing the light curve down to a timescale of $\sim$1.2 hours. For the analysis of the spot-only light curve, we binned the data every $10$ points, that is, we have one data point every $1200$\,s.

To compute the first 40 orbital harmonics we did a linear least squares fit to the entire TESS data train of the form:
\begin{equation}
L(t) = A + \sum_{\rm n =1}^{N} B_n \sin[2 n \pi (t-t_0)/P_{\rm orb}] + \sum_{n=1}^{N} C_n \cos[2 n \pi (t-t_0)/P_{\rm orb}]\,.
\end{equation}
In all, there are $2N+1$ free linear parameters, since $t_0$ can just be the time of the first data point, and $P_{\rm orb}$ is the orbital period determined separately as described above. For $N = 40$ this required inverting an $81 \times 81$ matrix, which takes just seconds on any modern laptop. 

Once the fit was completed, the 81 parameters could be used to evaluate the model light curve at any time and for any cadence. As an example, Fig.\,\ref{fig:example} shows the reconstructed orbital light curve of V521\,Dra and, after subtracting it from a small part of the raw TESS data, the residual variations caused by stellar activity. The brightness is expressed in magnitudes using TESS magnitude of 9.027 from TIC v8.2. We have repeated this exercise using 80 and 160 orbital harmonics with no discernible difference. We note that the fit did not remove the signals from spot eclipses during eclipses, but rather left an averaged out signal of the eclipse, including spot eclipses. For this reason, it would not be suitable as a baseline for eclipse mapping.

\section{Flares \label{app:flares}}

\begin{figure}[t!]
    \includegraphics[width=\columnwidth]{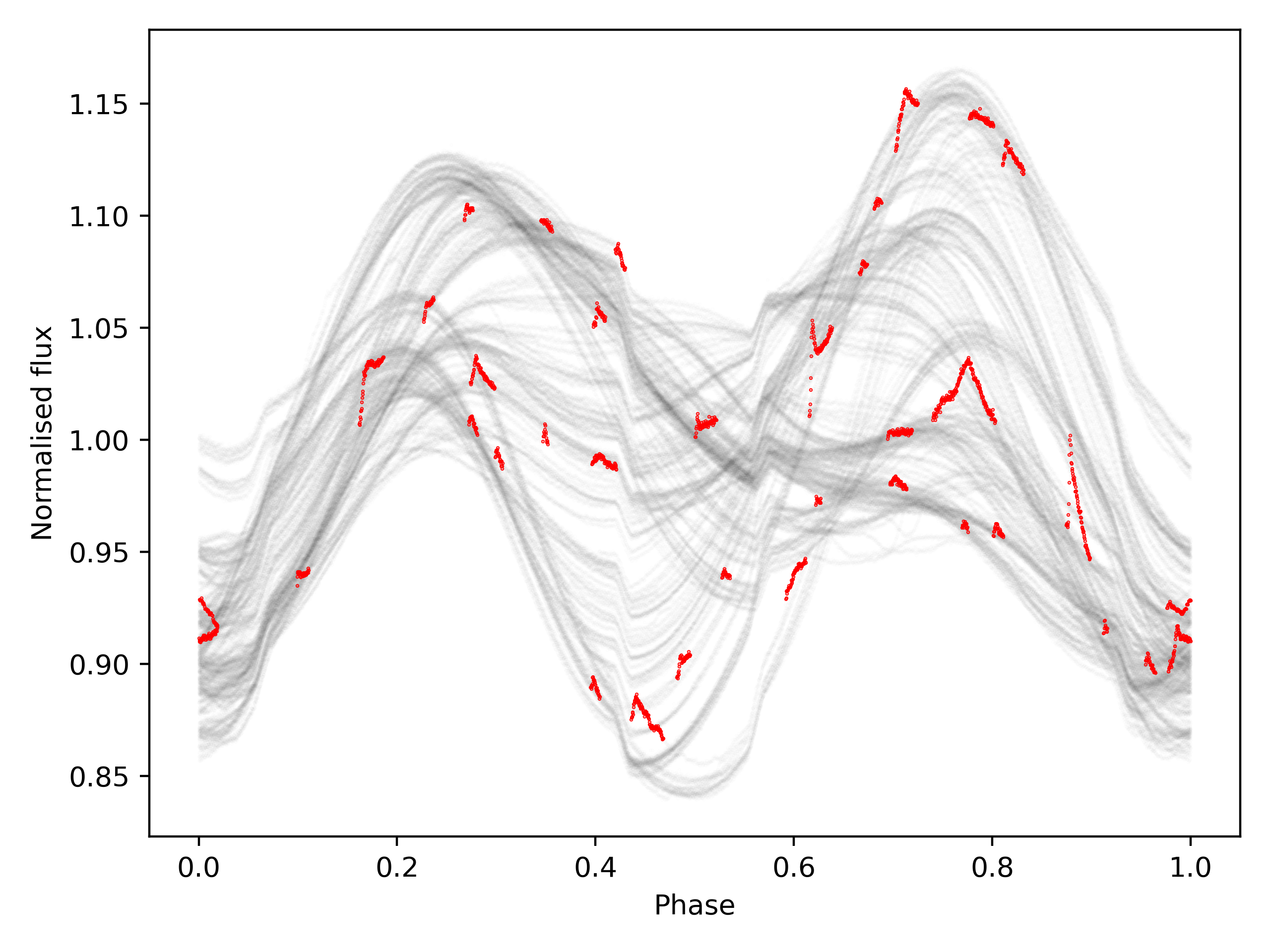}
      \caption{Phase folded light curve of V521\,Dra using all TESS sectors. Flares are marked with red.}
      \label{fig:flares}
\end{figure}

We manually identified 35 flares in 40 sectors of TESS photometry. These flares are displayed in the complete, phase folded TESS time series (Fig.\,\ref{fig:flares}) The flares show no significant phase preference, as indicated by Kuiper's test \citep{kuiper1960tests} that could not reject the uniform phase distribution with $p=0.26$ (see \citealt{2023A&A...674A.143K} for a similar treatment). Flare energies were calculated with the same method as in \citet{2025A&A...694A.161S}, they are in the range of $10^{34} - 10^{36}$\,erg in the TESS photometric band. The flare frequency distribution power-law index is $\alpha_{\rm flare} = 1.92 \pm 0.03$. These values are typical for red giant stars, see \citet{2021A&A...647A..62O}.

\section{Radial velocity  extraction \label{app:rv_analysis}}

We employed the broadening function (BF) formalism \citep{rucinski1999} with synthetic templates from \citet{coelho2005}, and conducted the analysis in two separate stages for the two components. For V521\,Dra~A, the average BF was derived over the wavelength range 4500$-$5825\,Å to avoid regions contaminated by telluric and interstellar Na lines. Singular-value decomposition of the synthetic template spectrum was used to construct the BF, followed by Gaussian smoothing with $\sigma = 2$\,km\,s$^{-1}$ to suppress regression noise. A rotational broadening model was then fitted using non-linear least squares, with uncertainties on the BF estimated from continuum regions. Linear limb darkening was fixed using an interpolated coefficient from \citet{claret2000}, and RVs were extracted from the best-fit models. To assess the impact of systematics, we divided each spectrum into five sub-intervals and repeated the RV fit. The final RVs were computed as the unweighted mean and standard deviation of these sub-sample fits.

We then co-added the spectra in the rest frame of component~A to obtain its dominant signal, and subtracted it from the individual spectra to search for the fainter companion in the residual. A narrow secondary signal was recovered in five of the spectra, consistent with a slowly rotating star. These detections were modelled using Gaussian profiles with parameters for RV, amplitude, width, and continuum level. 

We modelled the RV measurements with two Keplerian orbits shifted by a reference velocity $\gamma$. The RV curve of each component is fully described by five orbital parameters: the RV semi-amplitude $K$, the argument of periastron $\omega$, the eccentricity $e$, the mean anomaly $M_0$, and the orbital period $P_{\rm orb}$. We used a joint model for the two components, with a common eccentricity and coupled mean anomalies and arguments of periastron. We fixed the value of $P_{\rm orb}$ to that obtained from the analysis of eclipse light curves.

We used the MCMC algorithm implemented in \texttt{emcee} \citep{foreman2013} to sample a joint RV-model for the two components. Two Gaussian noise terms were used in the sampling for the two components, to account for astrophysical noise (varying line profile shape) in the RV observations. We employed eight~walkers per parameter for $50\,000$~iterations in the samplings.

The sampling yielded eccentricities consistent with zero and broad, poorly constrained posterior distributions for the periodic parameters $\omega_{\rm A}$ and $M_{\rm 0,A}$. When eccentricity is statistically indistinguishable from zero, these two parameters become degenerate and effectively describe only the orbital phase, leading to strong mutual correlations and large formal uncertainties. To test the significance of the model with eccentric orbits, we compared them with circular orbits. For this, we performed a second round of samplings with eccentric orbits, where the eccentricity was parametrised as $\ln e$. This lets the eccentricity take arbitrarily small values, while enforcing that it remains positive. We calculated the Bayes factor for the two models based on Bayesian information criterion (BIC) estimates \citep{liddle2007}. We found that the Bayes factor of an eccentric orbit in favour of a circular orbit is $B_{1,0} = 1.13$, which corresponds to insignificant evidence on the scale of \citet{jeffreys1961} and \citet{kass1995}.

Convective blueshift \citep{gray2009} or gravitational redshift \citep{einstein1952} may have a significant impact relative to the uncertainties of the RV data. This could in principle require fitting separate systemic velocities for the two components. We resampled the RV data allowing for independent system velocities for both components and compared this model to one with a single shared systemic velocity using the Bayes factor. We  found no evidence to support the choice of different radial velocity offsets for the two components.

In Table\,\ref{table:radvelAB}, we present the radial velocity values for components A and B of V521\,Dra taken with the FIES spectrograph.

\begin{table}[htb!]
\caption{Radial velocities for both components.
}
\label{table:radvelAB}
\centering
\begin{tabular}{c c r r}
\hline\hline
Phase & BJD & RV$_A$ & RV$_B$ \\
 & & (km\,s$^{-1}$) & (km\,s$^{-1}$)\\
\hline
0.0235 & 2460547.49149 & $ -31.0 \pm 0.6$ & \\ 
0.0797 & 2460560.54249 & $ -51.4 \pm 0.5$ & \\ 
0.2861 & 2460548.61270 & $ -79.3 \pm 0.9$ & $  96.2 \pm 1.6$ \\ 
0.3656 & 2460574.57421 & $ -65.6 \pm 0.7$ & $  65.6 \pm 1.3$ \\ 
0.5221 & 2460549.62056 & $ -15.3 \pm 0.8$ & \\ 
0.6587 & 2460554.47424 & $  18.9 \pm 0.6$ & $-125.1 \pm 1.1$ \\ 
0.7309 & 2460550.51230 & $  26.1 \pm 1.3$ & $-148.2 \pm 1.3$ \\ 
0.8843 & 2460585.32999 & $  10.5 \pm 0.6$ & $-108.3 \pm 1.1$ \\
\hline 
\end{tabular}
\end{table}

\section{Ca~II H \& K emission}\label{app:caHK}

\begin{figure}[htb!]
    \includegraphics[width=\columnwidth]{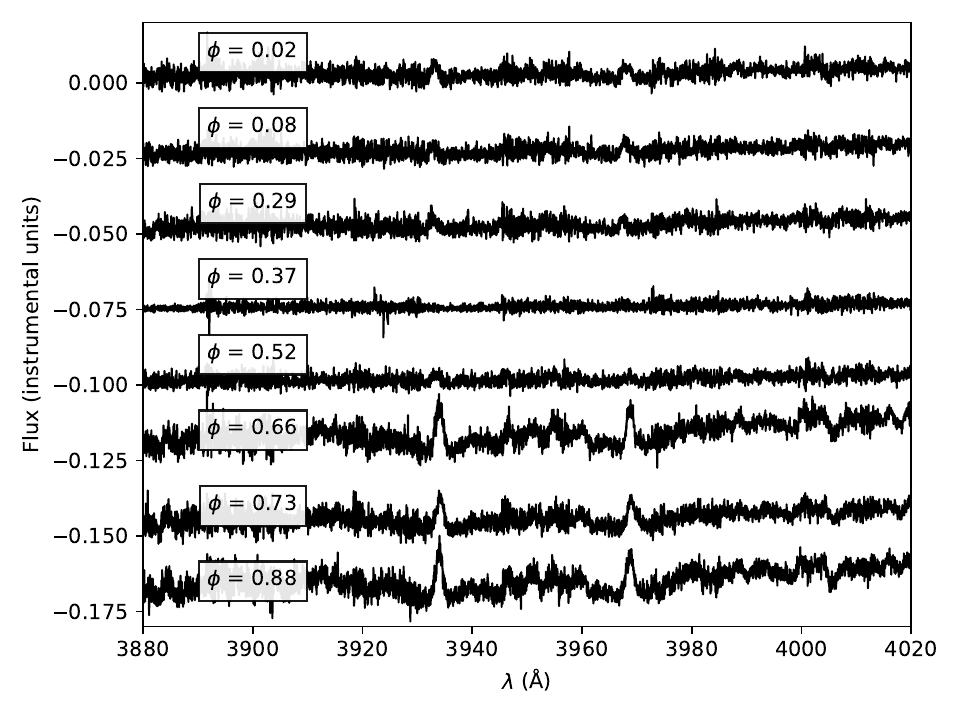}
      \caption{Orbital phase ($\phi$) ordered spectra of V521\,Dra around the Ca~II H\&K emission lines, obtained with the FIES spectrograph at the NOT. $\phi=0$ corresponds to the mid-time of the primary eclipse.}
      \label{fig:caHK}
\end{figure}

We quantified the chromospheric emission levels of V521\,Dra~A by measuring the Mount Wilson S-index \citep{vaughan1978}, and transforming it into the fractional emission flux $\log R'_{\rm HK}$ at the Ca~II H\&K line cores. For this transformation, we followed the procedure of \citet{mittag2013}, who detail the calculation of $\log R'_{\rm HK}$ for giants. We found periodic variations in the strength of the Ca~II H\&K emission. We obtained a maximum $S=1.91$ and $\log R_{\rm HK}' = -3.9$. Since the spectroscopic observations were taken over a span of 38 days, some development of the active regions is expected. For this reason, the emission strength does not strictly follow the orbital phase. The spectra around the Ca~II H\&K lines obtained with the FIES spectrograph are shown in Fig.\,\ref{fig:caHK}.

\section{Details of the SED modelling \label{app:sed}}

We fitted the composite SED of the binary in combination with the measured $K$ radial velocity semi-amplitudes and the relative quantities which were measured in our light curve analysis. 
Since the SED-fit was carried out with an MCMC analysis, the uncertainties are naturally propagated during the course of the fit. The goal, in short, was to fit the archival spectral flux measurements (SED points), the two spectroscopically determined $K$ RV semi-amplitudes, and the light curve-measured values of $i$, $a/R_{\rm A}$, $k_r$, and $k_f$, along with all their uncertainties to find the following eight parameters: the masses, radii, and effective temperatures of the two components, the distance, and the extinction to the source, $A_V$.  

To reduce the number of sampled parameters, we used the explicit assumption that the two stars in the binary have evolved coevally (i.e.~with no prior episode of mass transfer). In that case we can use stellar evolution tracks to solve for the following five parameters: the masses of the components, the age, distance, and extinction to the source. Specifically, the fitting code utilises MIST evolution tracks \citep{choi16, dotter16, paxton11, paxton15, paxton19}. For any given mass, age, and metallicity, the evolution tracks provide the stellar radius and $T_{\rm eff}$. The stellar atmosphere models were taken from \citet{castelli03}. The basic parts of this code have been extensively tested on a number of multistellar systems, including binaries \citep{yakut2025}, triples \citep{rappaport22}, and even a sextuple star system \citep{powell21}.  

We collected the available archival SED points from 0.42 to 22.1\,$\mu$m. We accessed these data from VizieR SED \citep{ochsenbein00} which, in turn, utilises systematic sky coverage of such surveys such as Pan-STARRS \citep{chambers16}, SDSS \citep{gunn98}, 2MASS \citep{2MASS}, WISE \citep{cutri2013}, and GALEX \citep{bianchi17}. 

Because the metallicity of V521\,Dra was only weakly constrained from the spectroscopic analysis, we ran the SED with a range of metallicities from ${-0.4 < \rm [Fe/H]} < +0.4$. We found that the best fit was obtained for ${\rm [Fe/H]} = -0.3 \pm 0.15$, but $\chi^2$ varied only weakly with [Fe/H].

\onecolumn
\section{Spot evolution}\label{app:spotevolution}

\begin{figure}[!htb]
    \centering
    \begin{minipage}[t]{0.48\columnwidth}
        \centering
        \includegraphics[width=\linewidth]{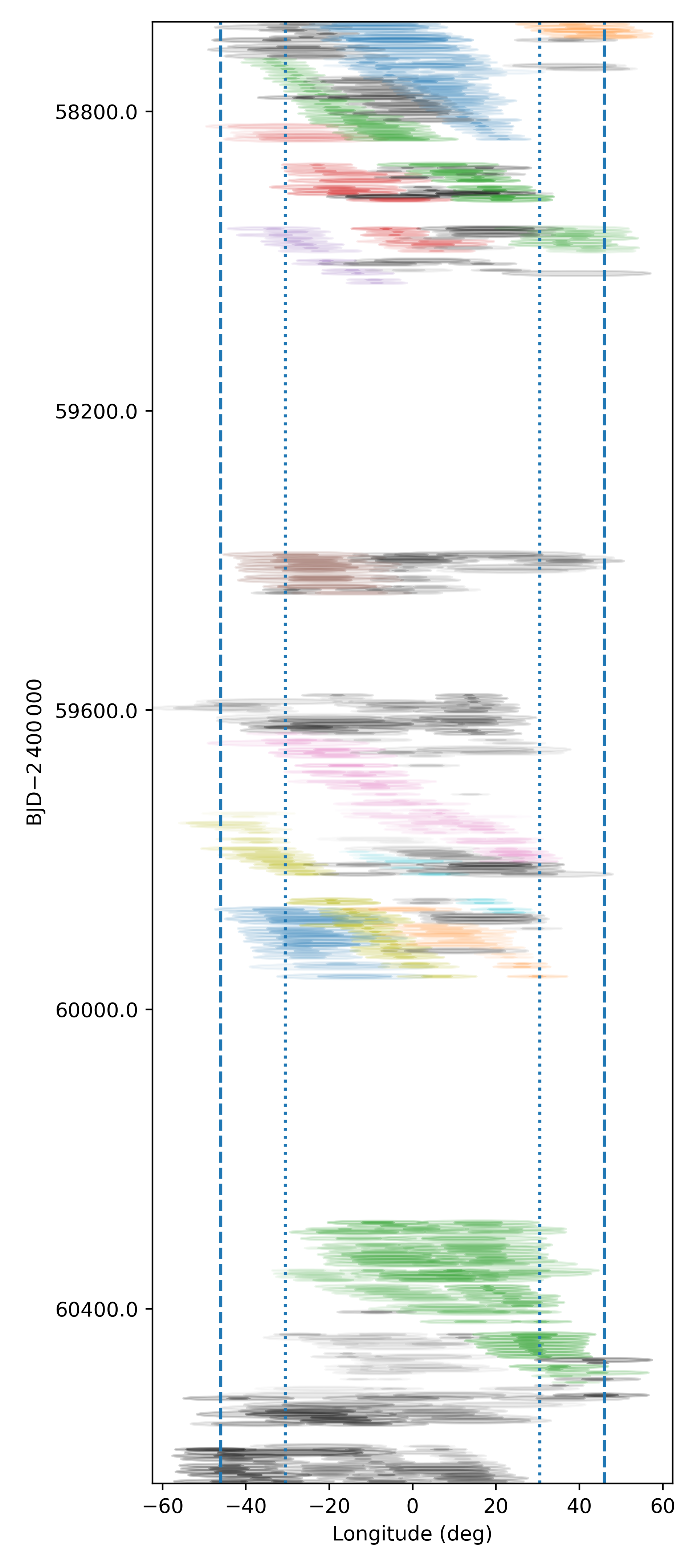}
        \caption{Coloured eclipse-mapping time series solutions.
        Dotted (dashed) lines correspond to $50\%$ ($10\%$)
        detectability limits.}
        \label{fig:eclipsemap_coloured}
    \end{minipage}\hfill
    \begin{minipage}[t]{0.48\columnwidth}
        \centering
        \includegraphics[width=\linewidth]{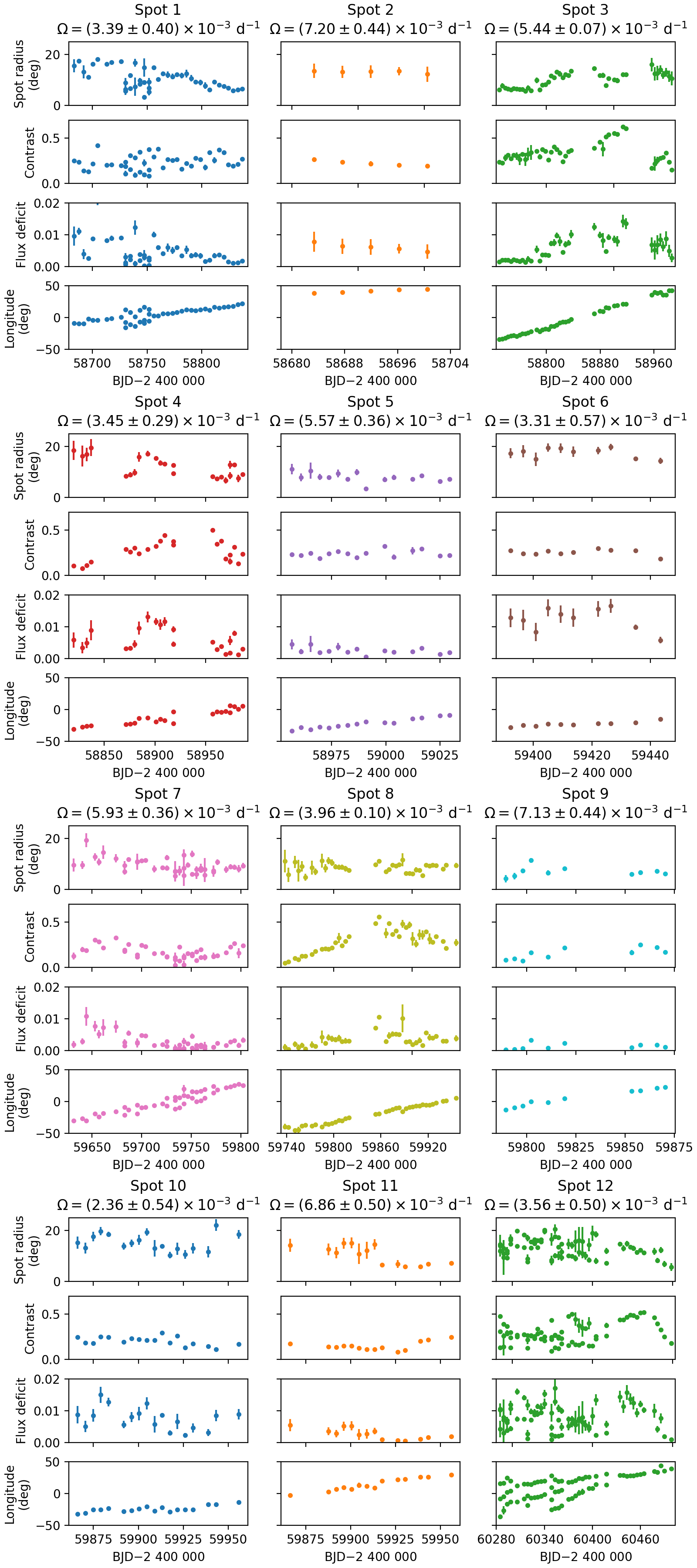}
        \caption{Evolution of spot properties for individual spot
        groups in the eclipse-mapping time series.}
        \label{fig:spotevolution_appendix}
    \end{minipage}
\end{figure}

In Fig.\,\ref{fig:eclipsemap_coloured}, we show the time series of the eclipse-mapping solutions, with confidently identified recurring spot groups marked with colour. We show the evolution of spot properties in Fig.\,\ref{fig:spotevolution_appendix}. The average angular velocities of the spot drifts in the orbital frame are given in the headers. The colour coding of individual spot groups is similar for both figures.

\end{appendix}

\end{document}